\documentclass[fleqn,usenatbib]{mnras}
\usepackage{newtxtext,newtxmath}
\usepackage[T1]{fontenc}

\DeclareRobustCommand{\VAN}[3]{#2}
\let\VANthebibliography\thebibliography
\def\thebibliography{\DeclareRobustCommand{\VAN}[3]{##3}\VANthebibliography}

\usepackage{graphicx}	% Including figure files
\usepackage{amsmath}	% Advanced maths commands
\usepackage{booktabs}
\usepackage{threeparttable}
\usepackage{orcidlink}

\newcommand{\pc}{\texttt{pop-cosmos}}
\newcommand{\mlim}{\textit{Ch.\,1}}
\definecolor{addred}{HTML}{CC0000}

\usepackage{algorithm}
\usepackage{algpseudocode}
\title[\pc: Star formation over 12 Gyr]{\pc: Star formation over 12 Gyr from generative modelling of a deep infrared-selected galaxy catalogue}

\author[S.\ Deger et al.]{Sinan Deger$^{\orcidlink{0000-0003-1943-723X}\,1}$\thanks{E-mail: sinan.deger@ast.cam.ac.uk},
Hiranya V.\ Peiris$^{\orcidlink{0000-0002-2519-584X}\,1,2}$,
Stephen Thorp$^{\orcidlink{0009-0005-6323-0457}\,1,2}$,
Daniel J.\ Mortlock$^{\orcidlink{0000-0002-0041-3783}\,3,4}$,
\newauthor
Gurjeet Jagwani$^{\orcidlink{0009-0004-7935-2785}\,1,5}$,
Justin Alsing$^{\orcidlink{0000-0003-4618-3546}\,2}$,
Boris Leistedt$^{\orcidlink{0000-0002-3962-9274}\,3}$,
and 
Joel Leja$^{\orcidlink{0000-0001-6755-1315}\,6,7,8}$
\\
$^{1}$ Institute of Astronomy and Kavli Institute for Cosmology, University of Cambridge, Madingley Road, Cambridge, CB3 0HA, UK\\
$^{2}$ The Oskar Klein Centre, Department of Physics, Stockholm University, AlbaNova University Centre, SE 106 91 Stockholm, Sweden\\
$^{3}$ Astrophysics Group, Imperial College London, Blackett Laboratory, Prince Consort Road, London, SW7 2AZ, UK\\
$^{4}$ Department of Mathematics, Imperial College London, London, SW7 2AZ, UK\\
$^{5}$ Research Computing Services, University of Cambridge, Roger Needham Building, 7 JJ Thomson Ave, Cambridge CB3 0RB, UK \\
$^{6}$ Department of Astronomy \& Astrophysics, The Pennsylvania State University, University Park, PA 16802, USA\\
$^{7}$ Institute for Computational \& Data Sciences, The Pennsylvania State University, University Park, PA 16802, USA\\
$^{8}$ Institute for Gravitation \& the Cosmos, The Pennsylvania State University, University Park, PA 16802, USA
}

\date{Accepted XXX. Received YYY; in original form ZZZ}

\pubyear{2025}

\begin{document}
\label{firstpage}
\pagerange{\pageref{firstpage}--\pageref{lastpage}}
\maketitle

% Abstract of the paper, 250 word limit
\begin{abstract}
We study star formation over $\sim$12~Gyr using \pc, a generative model trained on 26-band photometry of $\sim$420,000 COSMOS2020 galaxies (\textit{Spitzer} IRAC $\mlim < 26$). The model learns distributions over 16 stellar population synthesis parameters via score-based diffusion, matching observed colours and magnitudes. We use \pc\ to compute the cosmic star formation rate density (SFRD) to $z=3.5$ by directly integrating individual galaxy SFRs. The SFRD peaks at $z=1.3 \pm 0.1$, $\Delta z \simeq 0.6$ later than previous canonical estimates, with peak value $0.08\pm0.01$ $\mathrm{M}_{\odot}\,\text{yr}^{-1}\,\text{Mpc}^{-3}$. We classify star-forming (SF) and quiescent (Q) galaxies using specific SFR $<10^{-11}$ yr$^{-1}$, comparing with $NUVrJ$ colour selection.  The sSFR criterion yields up to 20 percent smaller quiescent fractions across $0<z<3.5$, with $NUVrJ$-selected samples contaminated by galaxies with sSFR up to $10^{-9}$ yr$^{-1}$. Our sSFR-selected stellar mass function shows a negligible number density of low-mass ($\lesssim10^{9.5}\,\mathrm{M}_\odot$) Q galaxies at $z\simeq1$, where colour-selection shows a prominent increase. Non-parametric star formation histories around the SFRD peak reveal distinct patterns: SF galaxies show gradually weakening correlations between their recent and earlier SFRs, implying increasingly stochastic star formation toward early epochs. Q galaxies exhibit full correlation ($r>0.95$) during the most recent $\sim$300 Myr, then sharp decorrelation with earlier star-forming epochs, marking clear quenching transitions. Massive ($10<\log_{10}(M_*/\mathrm{M}_{\odot})<11$) galaxies quench on a time-scale of $\sim1$~Gyr, with mass assembly concentrated in their first 3.5~Gyr. Finally, AGN activity (infrared torus luminosity fraction) peaks as massive ($\sim10^{10.5}\,\mathrm{M}_\odot$) galaxies approach the transition between star-forming and quiescent states, declining sharply once quiescence is established. This provides evidence that AGN feedback operates in a critical regime during the $\sim 1$ Gyr quenching transition.
\end{abstract}


%%%%%%%%%%%%%%%%%%%%%%%%%%%%%%%%%%%%%%%%%%%%%%%%%%

%%%%%%%%%%%%%%%%% BODY OF PAPER %%%%%%%%%%%%%%%%%%

\section{Introduction}
\label{sec:intro}
 
The global assembly of stellar mass in the Universe is linked to the star formation histories (SFHs) of individual galaxies. However, the SFH of any individual galaxy is difficult to constrain as neither photometric data nor spectra are sufficient to distinguish between the variety of possible gas accretion, feedback, and environmental processes \citep[see, e.g.][]{naab_ostriker17}. Effective SFH modelling hence needs to combine the flexibility to encode all the distinct models consistent with the data and the restrictions of a physically-motivated galaxy population prior. Inference of galaxy SFHs from observational data is typically carried out via spectral energy distribution (SED) fitting \citep[for a recent review see][]{pacifici23}. Two main approaches dominate SFH modelling: parametric and non-parametric.

Parametric methods assume physically-motivated functional forms (e.g.\ exponentially declining, double power-law, or lognormal; see examples in \citealp{buat08, dacunha08, maraston10, gladders13, simha14, abramson15, cisela17, diemer17, carnall18, carnall19}) but risk missing the diversity of actual galaxy SFHs \citep{leja19, lower20}. 

Non-parametric methods divide SFHs into time-bins with independent star formation rates \citep[see, e.g.][]{cid05, ocvirk06, tojiero07, kelson14, leja17, leja19_sfh}, providing flexibility to capture bursty or complex histories. However, without appropriate priors, non-parametric fits can become dominated by priors rather than signal. The robust inference of SFHs with non-parametric techniques relies heavily on the implementation of physically-informed priors \citep[see, e.g. \texttt{Prospector};][]{leja17, leja18, johnson21}, which requires great care since the inferred galaxy properties can be strongly dependent on the adopted prior \citep[see, e.g.][]{leja19_sfh, suess22, tacchella22a, tacchella22b, whitler23}\footnote{Parametric models are also prior-sensitive, and the assumption of a specific functional form can be viewed as a very strong implicit prior on the SFH \citep[see, e.g.][]{lee10, wuyts11, pforr12, carnall19, lower20, curtis21, sandles22, tacchella22a, tacchella22b}.}. \cite{leja19_sfh} has shown that $6$--$7$ time bins (logarithmically-spaced, following \citealp{ocvirk06}) optimally balance flexibility and information content if only photometric data is available. 

Recent results \citep{looser25, endsley25, witten25, wang25} from the \textit{James Webb Space Telescope} \citep[\textit{JWST};][]{jwst_mission} have reignited interest in understanding and empirically measuring SFH variability across cosmic time. Recent developments include sophisticated non-parametric methodologies such as the Dense Basis approach \citep{iyer17, iyer19}, spatially-resolved SFH models \citep{jain24, mosleh25}, detailed analysis of cosmological simulations \citep{iyer20}, and stochastic priors \citep{caplar19, tacchella20, wang20, iyer24, wan24} based on the power spectral density coupled with physical constraints imposed by the gas regulator model \citep{lilly13}. Individual SED-fitting of galaxies in large samples has become popular in SFH studies \citep{marchesini09, davidzon13, ilbert13, davidzon17, leja17, leja19_sfh, bellstedt20}, complementary to standard SFH variability measures such as the H$\alpha$-to-ultraviolet flux ratio \citep{weisz12, johnson13, sparre17, caplar19, faisst19, flores21}. However, it has become clear that SED-fitting of individual galaxies cannot robustly distinguish between bursty and smooth SFH \citep{wang24b}, motivating the development of realistic population-level priors \citep{wang25}.

We have developed a new approach for setting realistic, data-driven priors over the non-parametric seven-bin \citep{leja19_sfh} SFH description. The basis for this prior is our \pc\ generative model for the redshift-evolving galaxy population \citep{alsing20, alsing23, leistedt23, alsing24, thorp24, thorp25b, thorp25a}. The model is calibrated on $\sim\!420,000$ galaxies from the Cosmic Evolution Survey's \citep[COSMOS;][]{scoville07} COSMOS2020 \citep{weaver22} catalogue, subject to a \textit{Spitzer} IRAC $\mlim<26$ magnitude limit. This large sample therefore captures the full diversity of galaxy properties above the flux limit, including rare galaxies. The catalogue includes 26-band photometric data spanning the full ultraviolet (UV) to mid-infrared (MIR) wavelength range, benefitting from homogeneous depth, and self-consistent model-based photometric extraction \citep{weaver23_farmer}. 
The broad wavelength coverage of the COSMOS2020 catalogue, along with the inclusion of narrow and wideband data, makes it richly informative for \pc's physical description of galaxies in terms of a 16-parameter stellar population synthesis (SPS; for a review see \citealp{tinsley80, conroy13, iyer25}) model. 

By training the model on a single, clean photometric sample, and incorporating selection and data models (models of the measurement uncertainties) into the training process, \pc\ sidesteps many of the difficulties associated with incorporating auxiliary estimates of scaling relations or physical properties into the calibration of the model (as is done by empirical models; e.g.\ \citealp{moster18, behroozi19}), and obviates the need for calibrating the systematics and selection effects of these heterogeneous external analyses. The \pc\ model has been shown to reproduce key galaxy evolution scaling relations such as the stellar mass function \citep[SMF; see, e.g.][]{leja20, thorne21, driver22, weaver23, shuntov24, zalesky25}, star-forming main sequence \citep[see, e.g.][]{noeske07, daddi07, speagle14, leja22, sandles22, popesso23, fu24}, mass--metallicity relation \citep[see, e.g.][]{tremonti04, gallazzi05, gallazzi14, zahid17, cullen19, calabro21, kashino22, chartab24}, and fundamental metallicity relation \citep[see, e.g.][]{mannucci10, lara10, zahid14, cresci19, curti20, curti24}.

In this paper we utilize our \pc\ model to investigate the SFH of the galaxy population over cosmic time. In Section~\ref{sec:model_data} we summarize the \pc\ model and the resultant mock galaxy catalogues we use as the basis for our analysis. In Section~\ref{sec:sfrd} we use these mock catalogues to derive the evolving cosmic star formation rate density (SFRD, \citealt{madau14}) between $0<z<3.5$ by direct integration of synthetic galaxy SFRs (rather than using luminosity functions). In Section~\ref{sec:sf_q} we then look at the evolving fractions of star-forming (SF) and quiescent (Q) galaxies as identified using both colour-based selection \citep[e.g.][]{daddi04, wuyts07, williams09, ilbert10, ilbert13, arnouts13, leja19_uvj} and specific star formation rate (sSFR) \citep[e.g.][]{ilbert10, ilbert13, dominguez11}. We then investigate the different evolution of the SF and Q populations, looking at both their SMFs (Section~\ref{sec:smf}) and SFHs (Section~\ref{sec:sfh}). We discuss our results in Section~\ref{sec:discussion} and present our conclusions in Section~\ref{sec:conclusions}.

Cosmology-dependent quantities throughout the paper are computed assuming flat $\Lambda$CDM with $H_0=67.66$~km\,s$^{-1}$\,Mpc$^{-1}$ and $\Omega_{\textrm{m}}=0.3097$ \citep{planck18}.

\section{Data-Driven Model for the Galaxy Population}
\label{sec:model_data}

This analysis of the cosmic SFH is based on our \pc\ model for the redshift-evolving galaxy population \citep{alsing24, thorp25b}. This is a generative model -- i.e., it describes the data-generating process for galaxy photometry, computing SEDs from a population distribution of intrinsic galaxy properties, and adding observational effects such as noise and selection -- calibrated on a 26-band galaxy catalogue from COSMOS2020 \citep{weaver22} subject to a \textit{Spitzer} IRAC $\mlim<26$ selection. The training process accounts for the noise properties and data selection, such that the generative model describes the underlying galaxy population in the Universe, within the estimated mass and redshift completeness of the model. 

We use as input to our analysis our publicly available mock catalogue of 2 million model galaxies drawn from the trained \pc\ model. We give an overview of the generative model in Section \ref{subsec:model}, and describe auxiliary inputs based on SED fitting of COSMOS2020 galaxies in Section \ref{subsec:mcmc}. We introduce new rest-frame photometry emulators in Section \ref{subsec:restframe}, introduce the mock catalogue that we use in Section \ref{subsec:data}, and describe our volumetric normalization of mock galaxy counts in Section \ref{subsec:norm}.

\subsection{The \pc~galaxy population model}
\label{subsec:model}

We introduced the \pc\ model in \citet{alsing24}, where we calibrated the model based on an optically-selected ($r<25$) catalogue of $\sim140,000$ galaxies from COSMOS2020 \citep{weaver22}. In \citet{thorp25b} we updated the \pc\ model by re-training it on an MIR-selected ($\mlim<26$) catalogue of $\sim420,000$ galaxies from COSMOS2020, with this deeper selection expected to capture a larger fraction of the $z\lesssim6$ galaxy population with a high degree of completeness \citep[see][]{weaver23} when compared to deeper data in a smaller sub-field \citep[the Cosmic Assembly Near-infrared Deep Extragalactic Legacy Survey;][]{koekemoer11, grogin11, nayyeri17}. To enable successful modelling of this deeper data, \citet{thorp25b} made several improvements to the generative model, introducing a more flexible model for the distribution of photometric uncertainties in COSMOS2020, and improving the handling of low signal-to-noise (S/N) data \citep[following][]{lupton99}. 

The trained \pc\ generative model provides a complete recipe for generating mock photometric observations of the galaxy population out to $z<6$. It has four key elements:
\begin{enumerate}
    \item a population distribution over physical parameters;
    \item an emulated physical model that maps between these parameters and noiseless model fluxes;
    \item an uncertainty model that represents the distribution of flux uncertainties expected for a given survey, conditional on true flux;
    \item an error model that adds noise to the model fluxes to generate mock observations.
\end{enumerate}

The physical SPS model is based on the Flexible Stellar Population Synthesis \citep[\texttt{FSPS};][]{conroy09, conroy10a, conroy10b} and \texttt{Prospector} \citep{johnson21} frameworks. Specifically, we use a 16-parameter SPS model for galaxy SEDs, based on the \texttt{Prospector}-$\alpha$ parametrization developed by \citet{leja17, leja18, leja19_sfh, leja19}. We use: a \citet{chabrier03} stellar initial mass function (IMF); stellar libraries from the Medium-resolution Isaac Newton Telescope library of empirical spectra \citep[MILES;][]{sanchez06, falcon11}; the Modules for Experiments in Stellar Astrophysics (\texttt{MESA}; \citealp{paxton11, paxton13, paxton15, paxton18, paxton19}) Isochrones and Stellar Tracks \citep[MIST;][]{dotter16, choi16}; the \citet{byler17} nebular emission model grid generated using \texttt{CLOUDY} \citep{ferland13}; the \citet{draine07} dust emission templates; the \citet{nenkova08i, nenkova08ii} templates for hot dust emission from active galactic nuclei (AGNs); and the \citet{madau95} model for attenuation by the intergalactic medium (IGM). The 16 SPS parameters that characterize a galaxy are listed in Table \ref{tab:sps_parameters}. The key quantities are redshift; stellar mass; stellar and gas-phase metallicity; gas ionization; a two-parameter model for the hot dust torus around AGN \citep{leja18}; a seven-bin non-parametric SFH \citep{leja19_sfh}; and a three-parameter dust attenuation treatment with birth-cloud (affecting stars younger than 10~Myr; \citealp{charlot00}) and diffuse components, and a \citet{calzetti00} attenuation law with free slope \citep{noll09} and UV bump strength tied to the slope \citep{kriek13}. The seven-bin SFH is represented by six parameters which are defined as the base 10 logarithms of the star formation rate (SFR) ratios between adjacent star-forming bins. We use the same binning spacing as \citet{leja19_sfh, leja19}, which defines the most recent two bins to be of 30~Myr and 70~Myr in duration. The earliest bin is defined to encompass 15 percent of the age of the universe at the galaxy redshift. The other bins are logarithmically spaced in look-back time.

\begin{table}
    \centering
    \caption{Summary of SPS parameters used in the \pc\ galaxy population model \citep{thorp25b}. The learned \pc\ diffusion model defines the population-level prior over these parameters, and encodes the non-linear correlations between parameters.}
    \label{tab:sps_parameters}
    \begin{tabular}{l r}
        \toprule
        symbol / unit & description\\
        \midrule
        & \emph{base parameters} \\
        \cmidrule(l){2-2}
        $\log_{10}(M^\text{form}_*/\mathrm{M}_\odot)$ & logarithm of stellar mass formed\\
        $\log_{10}(Z_*/\mathrm{Z}_\odot)$ & logarithm of stellar metallicity \\
        $\Delta\log_{10}(\text{SFR})_{\{2:7\}}$ & ratios of SFR between adjacent SFH bins \\
        $\tau_2/\text{mag}$ & diffuse dust optical depth \\
        $n$  & index for diffuse dust attenuation law \\
        $\tau_1/\tau_2$ & birth cloud dust optical depth relative to $\tau_2$ \\
        $\ln(f_\text{AGN})$ & logarithm of AGN luminosity fraction \\
        $\ln(\tau_\text{AGN})$ & logarithm of AGN torus optical depth \\
        $\log_{10}(Z_\text{gas}/\mathrm{Z}_\odot)$ & logarithm of gas-phase metallicity \\
        $\log_{10}(U_\text{gas})$ & logarithm of gas ionization, $\frac{\text{photon density}}{\text{H density}}$ \\
        $z$ & redshift \\
        \midrule
        & \emph{derived quantities} \\
        \cmidrule(l){2-2}
        $t_\text{age}/\text{Gyr}$ & mass-weighted age \\
        $\log_{10}(M_*/\mathrm{M}_\odot)$ & logarithm of stellar mass remaining\\
        $\log_{10}(\text{SFR}/\mathrm{M}_\odot\,\text{yr}^{-1})$ & logarithm of SFR\\
        $\log_{10}(\text{sSFR}/\text{yr}^{-1})$ & logarithm of specific SFR \\
        \bottomrule
    \end{tabular}
\end{table}

The population distribution over the 16 free SPS parameters in the \pc\ model is represented by a score-based diffusion model \citep{song21}, which provides sufficient flexibility to represent a high-dimensional probability  density of the necessary complexity. Such a model represents a complex density or distribution of interest as being a continuous learnable transform between a simple base density (often a multivariate Gaussian) and the target, with the transform encoded by a stochastic differential equation (SDE) representing a diffusion process (with an equivalent ordinary differential equation representation; \citealp{song21}). The learnable element of the model is an approximation to the `score' function, which describes the gradient of the density as it is transformed from the base density and target density (and vice-versa). Our diffusion model configuration is described in detail in \citet{alsing24} and \citet{thorp25b}, and is based on a variance-exploding SDE \citep{song19, song21}, closely following the framework described in \citet{song21}.

Draws of the SPS parameters from the trained diffusion model are transformed to noiseless 26-band COSMOS-like fluxes using an emulator for \texttt{FSPS} \citep[\texttt{Speculator};][]{alsing20} that was trained to predict \texttt{FSPS} photometry conditional on SPS parameters, as described in \citet{thorp25b}. To add realistic noise to the model fluxes, in \citet{thorp25b} we trained an uncertainty model to reproduce the distribution of flux uncertainty conditional on flux. This model was trained based on the catalogued fluxes and flux uncertainties reported by \citet{weaver22}. The 26-dimensional conditional distribution is represented by a second score-based diffusion model (with variance-preserving SDE; \citealp{sohl15, ho20, song21}), trained via denoising score-matching \citep{hyvarinen05, song19, song21}. Given these uncertainties, flux errors are added to the model fluxes using Student's $t$ distributions \citep[following][]{leistedt23}. 

During the training of the \pc\ model, the distributions of noisy model colours and magnitudes (in the asinh system introduced by \citealp{lupton99}) generated from the full forward process are compared to the COSMOS2020 colours and magnitudes, with the similarity between the two being assessed using a series of summary statistics \citep[defined in][]{thorp25b}. The diffusion model defining the population distribution over SPS parameters is adjusted (via stochastic gradient descent with \texttt{Adam}; \citealp{kingma14}) until the difference between model and data is minimized. The training of the diffusion model is carried out simultaneously with the fitting of a set of calibration parameters (introduced by \citealp{leistedt23}), allowing for band-by-band zero-point offsets, and line-by-line corrections to the strength and variance of nebular emission. The emission line corrections are applied on top of the standard \texttt{CLOUDY}-based nebular model grid from \citet{byler17} included in \texttt{FSPS}; the corrections tend to be small, and have remained fairly consistent between \citet{leistedt23}, \citet{alsing24}, and \citet{thorp25b}. The model (including the optimization of zero-point and emission line corrections) is trained using photometry alone, and is validated in colour and magnitude space \citep[see][]{thorp25a}, and by comparing its astrophysical predictions to literature estimates of well-known scaling relations.

\subsection{SED fits for COSMOS2020 with the \pc\ prior}
\label{subsec:mcmc}
In \citet{thorp24} we presented a method for performing Bayesian SED fits to individual galaxies, using the \pc\ model as the prior distribution for the SPS parameters. We developed a GPU-enabled workflow for this, using the \texttt{Speculator} emulator \citep{alsing20} for fast SPS model evaluation when performing Markov-chain Monte Carlo (MCMC) sampling of the posterior. We used an affine-invariant ensemble sampler \citep{goodman10} using the `parallel stretch' move \citep{foreman13} to enable effective vectorization on GPU hardware, implemented using \texttt{PyTorch} \citep{paszke19} in the \texttt{affine} package\footnote{\url{https://github.com/justinalsing/affine}}. In using the \pc\ diffusion model as a prior we exploited the `probability flow' interpretation of score-based models \citep[see][]{song21}, which enables deterministic evaluation of the probability density for any set of parameters by solving a neural ordinary differential equation \citep{chen18, grathwohl18}. In \citet{thorp24} we used this pipeline to obtain posterior samples of the 16 \pc\ SPS parameters for $\sim\!230,000$ COSMOS2020 galaxies, based on their 26-band photometry.

We then updated this analysis using the retrained \pc\ model presented in \cite{thorp25b} (and which is summarized in our Section~\ref{subsec:model}), obtaining posterior samples of the 16 SPS parameters for the full $\sim420,000$-galaxy $\mlim<26$ COSMOS2020 catalogue. In this work we use the \citet{thorp25b} posteriors on stellar mass, SFR, and redshift as secondary inputs to our analysis. These quantities will be used primarily in Section \ref{subsec:data} to set upper limits on stellar mass and SFR, and to estimate the $z<4$ number count in COSMOS2020 in Section \ref{subsec:norm}.

\subsection{Rest-frame photometry emulators for \pc}
\label{subsec:restframe}

Since the trained generative model is represented by an SPS parametrization, its predictions are not specific to the data on which it was calibrated. Draws of the SPS parameters from the \pc\ model can therefore be used to synthesize both full SEDs and photometry in passbands other than the 26 COSMOS bands used in the model training. For this work, we use  \texttt{Speculator} \citep{alsing20} to train a new set of neural emulators that predict rest-frame absolute magnitudes in the \textit{GALEX} $NUV$, HSC $r$, and UltraVISTA $J$ ($NUVrJ$) bands. We emulate photometry in these bands in particular as they are commonly used to define colour-based selections of star-forming and quiescent galaxies \citep[e.g.][]{wuyts07, williams09, ilbert13, leja19_uvj}. 

\subsection{The \pc\ mock galaxy catalogue}
\label{subsec:data}

While the \pc\ model of the galaxy population is in the form of a 16-dimensional distribution of SPS parameters, it is simplest to work with mock galaxy catalogues generated from this distribution. Each galaxy drawn from \pc\ which is used in this work has: individual values for (i) the 16 SPS parameters; (ii) the key derived quantities listed in Table~\ref{tab:sps_parameters}; (iii) both noiseless and noisy observer-frame fluxes in the 26 COSMOS passbands; and (iv) noiseless rest-frame absolute magnitudes in the $NUVrJ$ bands.  Where we quote (s)SFRs, these are defined as being the average over the last 100~Myr. We adopt a 100~Myr timescale as this is widely used or implicit in calibrations of observational SFR indicators \citep[see e.g.][]{kennicutt98, kennicutt12}, including colour-based tracers informed by the UV slope of the SED \citep[see e.g.][]{arnouts13}. It can also be directly computed from the two most recent bins of our 7-bin non-parametric SFH \citep{leja19_sfh, leja19}.

In this paper we use the $\mlim<26$ mock catalogue of 2 million galaxies presented by \citet{thorp25b}, augmented with the rest-frame $NUVrJ$ predictions listed in Section~\ref{subsec:restframe}. From this we select only galaxies that are more massive than the redshift-dependent stellar mass completeness limit given in \citet{thorp25b}. As in \citet{alsing24}, this limit is defined based on the turnover of the SMF, i.e.\ the mode at a given redshift. This completeness limit is conservative compared to the COSMOS2020 completeness limit given in \citet{weaver23}, around $0.1$--$0.3$~dex higher in stellar mass at all redshifts $\gtrsim0.3$. Additionally, we select only those galaxies with $z<4$, as the COSMOS2020 redshift distribution declines significantly beyond this. Finally, we impose a physically-motivated upper limit on the galaxies' SFR based on the most extreme stellar masses and SFRs identified in SED fits (see Section~\ref{subsec:mcmc}) to the COSMOS2020 galaxies \citep{thorp25b}\footnote{Specifically, for each galaxy we take the $97.5$th percentile of its posterior as an upper limit on both its stellar mass or SFR. We then compute a histogram of these individual upper limits for the COSMOS2020 catalogue, and take the $99.9$th percentile of this distribution. This gives an estimate of the most extreme values plausibly probed by COSMOS2020, yielding upper limits of stellar mass $<10^{11.6}\,\mathrm{M}_\odot$, and $\text{SFR}<10^{2.8}\,\mathrm{M}_\odot\,\text{yr}^{-1}$. {These upper limits are constant at all redshifts, unlike the mass completeness limit which is redshift-dependent. The upper limit on SFR aligns reasonably well with the highest typical SFRs estimated from radio and/or far IR indicators at $z\lesssim3$ \citep[e.g.][]{gruppioni13, schreiber15, tomczak16, molnar18, leslie20}, including those of ultra-luminous IR galaxies with $10^2\lesssim\text{SFR}/\mathrm{M}_\odot\,\text{yr}^{-1}\lesssim10^3$, but perhaps truncating some rare hyper-luminous IR galaxies ($\text{SFR}\gtrsim10^3\mathrm{M}_\odot\,\text{yr}^{-1}$) that do not significantly contribute to the total SFRD \citep[see e.g.][]{novak17}.}}. This sets realistic expectations as to what we can reliably learn from the COSMOS2020 data-set. Together, the redshift, SFR, and stellar mass cuts reduce the input catalogue of 2,000,000 galaxies to a sample of 1,331,800 galaxies which are used in the remainder of the paper. 

{All subsequent results in the paper (i.e., constraints on population level properties such as the SFRD, SMF, and quiescent fraction) are computed from the \pc\ mock galaxy catalogue, and \emph{not} from the per-galaxy posteriors of COSMOS2020 sources. Whilst finite in size, the mock catalogue is statistically representative of the trained \pc\ model, which is constrained by COSMOS2020 photometry at the population level as described in \citet{alsing24} and \citet{thorp25b}. In this sense, \pc\ and the subsequent analyses in this paper are closer in philosophy to the approach taken by empirical models \citep[e.g.][]{becker15, moster18, behroozi19, openuniverse25, alarcon25} than to analyses that combine collections of galaxy-level posteriors \citep[e.g.][]{leja20, leja22, lewis25}.} 

{However, \pc\ is distinct from these empirical models. The latter attempt to post-facto account for systematic shifts in galaxy evolution trends between compilations of literature studies, whereas the calibration of \pc\ involves direct forward modelling of observations, including noise properties and selection effects. Hence, \pc\ fully encompasses the statistical measurement uncertainties present in the data. There is no additional information available in derived posteriors on redshifts or stellar masses that have been conditioned on the same photometry.}

\subsection{Absolute normalization}
\label{subsec:norm}

By default, mock catalogues generated from \pc\ have arbitrary normalization, in the sense that there is no constraint on the number of galaxies generated.  Such catalogues can be used to explore the properties of individual galaxies or relative demographics, but an absolute normalization is required to estimate cosmological densities.  We hence normalize the catalogues by matching to the COSMOS2020 survey on which \pc\ is calibrated, using the observed numbers brighter than the $\mlim$ magnitude limit. Following \cite{weaver23}, we use the {`combined mask'} area of $\Omega = 1.27$~deg$^{2}$ (as opposed to the nominal survey area of 2 deg$^{2}$). {This is the subset of the COSMOS field with overlapping HSC, UltraVISTA, and \textit{Spitzer} IRAC coverage, with regions around bright stars and artifacts removed \citep{weaver22}.} We use the \cite{thorp25b} catalogue of inferred galaxy properties described in Section~\ref{subsec:mcmc} to determine the number of galaxies in this area out to our adopted maximum redshift of $z = 4$. We further restrict the sample to galaxies that satisfy the mass-completeness prescription described in \cite{thorp25b}.  This leaves $N = 332,950$ galaxies, implying a reference number density of $\Sigma = N / \Omega \simeq 2.60\times10^5$~deg$^{-2}$ for the $z < 4$ galaxy population satisfying our mass-completeness requirement\footnote{{This value of $N$ is computed based on posterior medians of stellar mass and $z$ from \citet{thorp25b}; this neglects the uncertainties on these parameters, which would propagate to a $\lesssim\!4$ percent redshift-independent systematic uncertainty in this overall normalization factor.}}.

\section{The Cosmic Star Formation Rate Density}
\label{sec:sfrd}

\begin{figure*}
    \centering
    \includegraphics[width=0.75\linewidth]{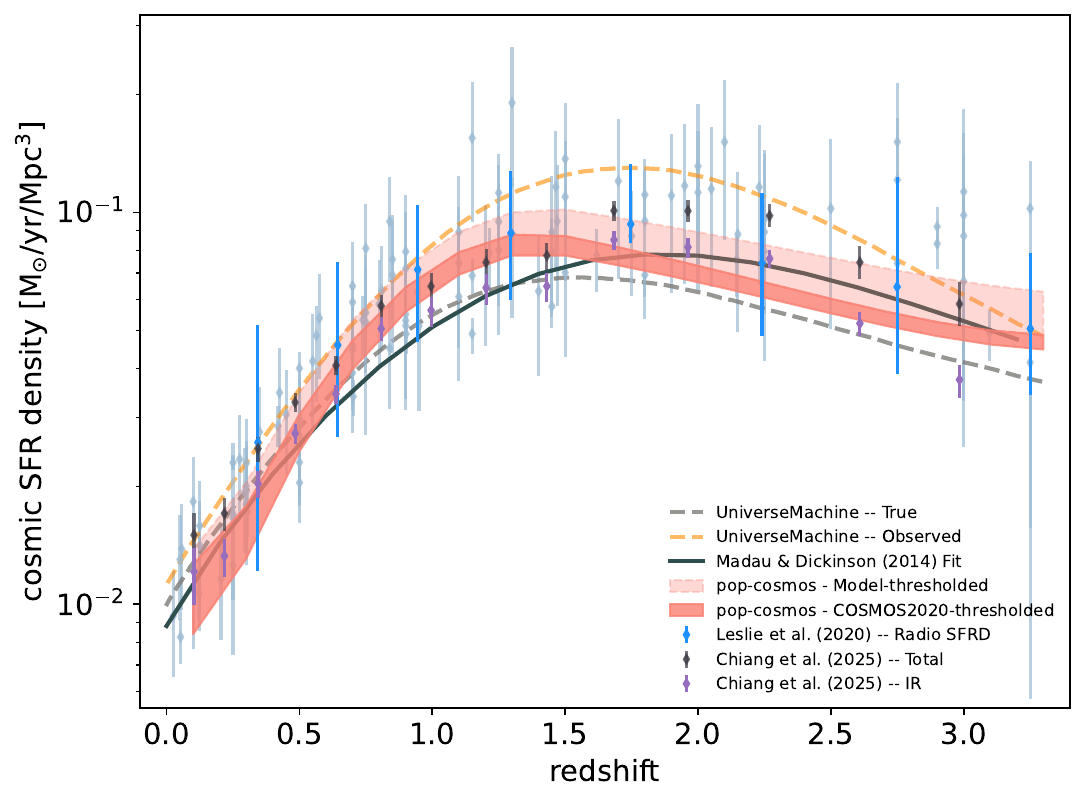}
    \caption{The \pc\ cosmic SFRD in bins of width $\Delta z=0.2$. Dark- and light-red shaded regions show, respectively, our COSMOS2020- and model-thresholded estimates. Black curve is from \citet{madau14}. Gray and orange curves are from  \citet{behroozi19}. Blue markers show the literature compilation assembled by \citet{behroozi19} and updated in this paper. Black and purple markers highlight results from the new CIB-based analysis of \citet{chiang25}, respectively with and without a UV-luminosity correction. {Solid blue markers highlight the 3~GHz radio SFRD from \citet{leslie20}.} All results are presented assuming a \citet{chabrier03} IMF.}
    \label{fig:sfrd}
\end{figure*}

The cosmic SFRD \citep{lilly96, madau96c} is the total mass formed in stars per unit time per unit comoving volume as a function of redshift, and is a key observable for understanding galaxy evolution \citep[for a review see][]{madau14}. The standard approach to measuring cosmic SFRD involves integrating UV or IR luminosity functions (LFs) across redshift bins, as described in the \citet{madau14} compilation. In this approach LFs are typically assumed to have a simple mathematical form, such as a \cite{schechter76} function, which is then fit to rest-frame UV (typically 1500 or 2800 \AA) or total IR (8--1000~$\upmu$m) luminosities. These LFs are then integrated down to a minimum luminosity limit to obtain the luminosity density. The conversion from luminosity density to SFRD requires calibration factors derived from SPS models, assuming a universal IMF and star formation time-scales. For UV-derived SFRDs, dust correction factors must also be applied, typically using estimates of the UV slope, $\beta$, or IR/UV ratios when available \citep[e.g.][]{dunne09, bouwens12}. 

An LF-based approach to SFRD estimation faces several fundamental limitations: (i) the assumed functional form is not flexible enough to capture the true shape of the LF particularly at the faint end \citep{lan16}; (ii) extrapolation below observational limits can introduce factors of $\sim2$ uncertainty in the integrated SFRD \citep{lilly96}; (iii) dust corrections remain highly uncertain at $z>2$ where IR observations are limited \citep{kobayashi13, mclure18_dust}; (iv) the conversion from luminosity to SFR assumes a constant star formation history over $\sim100$~Myr \citep[e.g.][]{madau14}, which fails for bursty or recently quenched galaxies; and (v) combining heterogeneous datasets across different redshift ranges and survey depths potentially introduces systematic uncertainties that are difficult to quantify \citep{madau14, driver18, behroozi19}. Previous work by \citet{leja19} has shown that self-consistent modelling of galaxy SFHs and stellar mass (e.g.\ with \texttt{Prospector}) is required to obtain non-conflicting mass growth and SFRD estimates \citep[see also][]{leja15}.

Here, we use direct integration of individual galaxy SFRs from our generative model to construct the SFRD, circumventing several of the above issues faced by LF-based methods, in particular by avoiding assumptions about the functional form of the LF. Complementary results, presenting the \pc\ estimate of the redshift-evolving star-forming sequence can be found in \citet{thorp25b}, who show that it agrees quantitively with the results of \citet{leja22} in the redshift range where the analyses overlap. The success of our method relies on accurate SFH recovery for large, complete samples; the data-driven calibration by \citet{thorp25b} of \pc\ to a deep IR-selected multi-wavelength catalogue of $\sim\!420,000$ galaxies from COSMOS2020 \citep{weaver22} meets this requirement. We describe in Section \ref{subsec:sfrd_method} how we use \pc\ to compute the cosmic SFRD, including uncertainty estimates, and then we present our results and compare them with results from the literature in Section \ref{subsec:sfrd_result}. 

\subsection{Methodology}
\label{subsec:sfrd_method}

Here we describe the methodology we use to go from the \pc\ model to redshift-binned estimates of the SFRD, $\Psi$, in Section \ref{subsec:estimate}, and the associated uncertainty, $\sigma_\Psi$, in Section \ref{subsec:sfrd_unc}.

\subsubsection{SFRD estimate}
\label{subsec:estimate}

We estimate the cosmic SFRD in $B$ redshift bins of width $\Delta z = 0.2$, with the $b$'th bin extending from $z_b - \Delta z / 2$ to $z_b + \Delta z / 2$ (for $b \in \{1, 2, \ldots, B\}$ and by default with $z_1 = 0.1$ and $z_B = 3.5$). In a flat universe the co-moving volume over the whole celestial sphere out to redshift $z$ is, from e.g.\ \cite{Hogg99}, $V_{\rm co}(z) = (4\pi / 3) \, D^{3}_{\rm co}(z)$,
where $D_{\rm co}(z)$ is the co-moving distance to redshift $z$.  The co-moving volume of the $b$'th bin within the COSMOS sky area $\Omega$ (Section~\ref{subsec:norm}) is hence 
\begin{equation}
    V_{{\rm co},b} = \frac{\Omega}{3} 
    \left[D^{3}_{\rm co}(z_b + \Delta z / 2) - D^{3}_{\rm co}(z_b - \Delta z / 2)\right].
\end{equation}

To calculate the SFRD in bin $b$ we draw a large sample of $N_b$ galaxies without replacement from our input catalogue (Section~\ref{subsec:data}), subject to the constraint that their redshifts are between $z_b - \Delta z / 2$ and $z_b + \Delta z / 2$. The acceptance fraction is then the \pc\ prediction for the fraction of galaxies in the bin, $f_b$. The \pc\ model provides the SFR of the $n$'th accepted galaxy as $\psi_n$ (with $n \in \{1, 2, \ldots, N_b\}$).  The cosmic SFRD at redshift $z_b$ is then estimated as
\begin{equation} \label{eq:sfrd}
    \Psi_{b} = \frac{N f_b}{V_{{\rm co},b}}
    \frac{1}{N_b} \sum_{n=1}^{N_b} \psi_n,
\end{equation}
where $N$ is the number of actual COSMOS2020 galaxies satisfying our selection cuts (from Section~\ref{subsec:norm}). Although not a formal requirement, the default is that $N_b > N f_b$ in order to reduce shot noise; choosing $N_b = N f_b$ would be the equivalent of working with the COSMOS2020 sample directly.

The SFRD estimate computed by summing SFRs as in Equation~\ref{eq:sfrd} is especially sensitive to extreme outliers in the galaxy population, since we do not regularize the form of the SFR distribution (e.g. by adopting restrictive functional forms as used in LF-based methods). The empirical upper SFR threshold described in Section \ref{subsec:data} regularizes the model predictions to lie within the distribution found in the COSMOS2020 catalogue; however, \pc\ does extrapolate beyond this empirical upper limit. To faithfully represent the predictions of the model including this regime, we present a second `extrapolated' estimate with an upper SFR threshold, which is derived as follows. First, we take the mock catalogue with the completeness limits imposed, except for the COSMOS2020-based SFR threshold. We divide the catalogue into 20 redshift bins each containing 5 percent of the population. We estimate the 99.9th percentile of the SFR distribution in each bin and adopt this as an upper threshold when computing the SFRD estimate in that bin. The uncertainty on this extrapolated estimate is computed in the same manner as for the COSMOS2020-thresholded estimate.

\subsubsection{SFRD uncertainty}
\label{subsec:sfrd_unc}

The estimate of the cosmic SFRD given in Equation~\ref{eq:sfrd} is subject to several distinct sources of uncertainty: 
(i) Poisson noise from the overall COSMOS normalization (i.e. effectively $N$); 
(ii) additional cosmic variance from galaxy clustering on the scale of the COSMOS volume;
(iii) the sampling of the population within a bin, particularly the high-SFR tail. We do not have a full model for the effective posterior distribution of the cosmic SFRD, so resort to a heuristic bootstrap procedure to estimate the uncertainty in each redshift bin. This procedure carefully accounts for counting uncertainties; we do not attempt to estimate the systematic uncertainty from the adopted physical models, e.g.\ the IMF.

For each redshift bin $b$ we produce $J = 10^4$ independent bootstrap estimates of the SFRD, following the procedure described in Section~\ref{subsec:estimate}, but with the normalization fixed to the COSMOS2020 numbers and sky area.  For iteration $j$ (with $j \in \{1, 2, \ldots, J\}$) the procedure is as follows:

\begin{enumerate}

\item 
Following the recipe described by \cite{moster11}, we use the redshift limits, $z_b - \Delta z / 2$ and $z_b + \Delta z / 2$, and COSMOS2020 sky area, $\Omega$, to calculate the cosmic variance in the bin, $\sigma^2_{V,b}$. This accounts for the possibility that large-scale correlations in the galaxy distribution result in the COSMOS2020 field being systematically under-dense or over-dense relative to the cosmic mean.

\item 
We take the total variance in the galaxy numbers in the bin to be $\sigma_b^2 = N f_b \, (1 + N f_b \, \sigma^2_{V,b})$, where $N$ is the total number of COSMOS2020 galaxies (Section~\ref{subsec:norm}) and $f_b$ is the fraction of these in the redshift bin. In contrast to Section~\ref{subsec:estimate}, the numbers are fixed to the actual COSMOS2020 values to appropriately account for the shot noise which is inevitably present in any density estimates from this finite sample.  That said, the cosmic variance dominates over the purely Poisson contribution in most bins, so this is not critical in practice.  

\item 
For this bootstrap realisation we draw the number of galaxies in the bin, $N_{b,j}$, from a normal distribution of mean $N f_b$ and variance $\sigma_b^2$, rounded to the nearest integer.  Strictly, this should be a draw from a scaled Poisson distribution, but the galaxy numbers are sufficiently high that the normal approximation is adequate.

\item 
As in Section~\ref{subsec:estimate}, $N_{b,j}$ draws are made from our input catalogue, using rejection sampling to select only galaxies with redshifts between $z_b - \Delta z / 2$ and $z_b + \Delta z / 2$.

\item 
The $j$'th bootstrap estimate for the SFRD, $\Psi_{b,j}$, is then calculated as in Equation~\ref{eq:sfrd}, but with $N_b \rightarrow N_{b,j}$ and the SFR values for each of the $N_{b,j}$ galaxies particular to this iteration.
    
\end{enumerate}

We then estimate the uncertainty on the SFRD in the bin, $\sigma_{\Psi,b}$, as the mean-subtracted variance of the bootstrap realisations according to
\begin{equation}
\sigma^2_{\Psi,b} 
  = \frac{1}{J} 
  \sum_{j = 1}^J \left( \Psi_{b,j} - \bar{\Psi}_b\right)^2,
\end{equation}
where $\bar{\Psi}_b = \sum_{j = 1}^J \Psi_{b,j} / J$ is the mean of the $J$ bootstrap estimates of the SFRD. While this recipe is somewhat heuristic it does explicitly incorporate Poisson shot noise, cosmic variance and the stochastic nature of the \pc\ galaxy draws.

\begin{table}
\centering
\caption{Summary of literature SFRD measurements used in Figure~\ref{fig:sfrd}. These data are taken directly from the published sources with their original integration limits.}
\begin{threeparttable}
\label{tab:sfrd_literature}
\begin{tabular}{lcr}
\toprule
publication & redshifts & type\tnote{a} \\
\midrule
\cite{bellstedt24} & 0.0--9.6 & forensic\tnote{b} \\
\cite{chiang25} & 0.0--4.0 & CIB\tnote{c} \\
{\cite{cochrane23}} & {0.0--4.0} & {radio\tnote{d}} \\
\cite{cucciati12}\tnote{e,f} & 0.0--5.0 & UV\tnote{g} \\
\cite{drake15}\tnote{d} & 0.6--1.5 & [O\,{\sc ii}]\tnote{h} \\
\cite{dsilva23} & 0.0--4.7 &  forensic\tnote{b} \\
\cite{dunne09}\tnote{e,f} & 0.0--4.0 & radio\tnote{i} \\
{\cite{enia22}} & {0.0--3.5} & {radio\tnote{i}} \\
\cite{gunawardhana13}\tnote{e,f} & 0.0--0.35 & H$\alpha$\tnote{j} \\
\cite{kajisawa10}\tnote{e,f} & 0.5--3.5 & UV/IR\tnote{k} \\
\cite{karim11}\tnote{e} & 0.2--3.0 & radio\tnote{i} \\
\cite{leborgne09}\tnote{e,f} & 0.0--5.0 & IR--mm\tnote{l} \\
{\cite{leslie20}} & {0.0--5.0} & {radio\tnote{m}} \\
\cite{ly11a}\tnote{e,f} & 0.8 & H$\alpha$\tnote{j} \\
\cite{ly11b}\tnote{e,f} & 1.0--3.0 & UV\tnote{g} \\
\cite{magnelli11}\tnote{e,f} & 1.3--2.3 & IR\tnote{n} \\
{\cite{novak17}} & {0.0--5.0} & {radio\tnote{m}} \\
\cite{robotham11}\tnote{e,f} & 0.0--0.1 & UV\tnote{g} \\
{\cite{rowan-robinson16}} & {0.0--6.0} & {FIR\tnote{o}} \\
\cite{rujopakarn10}\tnote{e,f} & 0.0--1.2 & FIR\tnote{p} \\
\cite{salim07}\tnote{e} & 0.0--0.2 & UV\tnote{g} \\
\cite{santini09}\tnote{e,f} & 0.3--2.5 & IR\tnote{n} \\
\cite{schreiber15}\tnote{e,f} & 0.0--4.0 & FIR\tnote{p} \\
\cite{shim09}\tnote{e,f} & 0.7--1.9 & H$\alpha$\tnote{j} \\
\cite{sobral14}\tnote{e} & 0.4--2.3 & H$\alpha$\tnote{j} \\
\cite{zheng07}\tnote{e} & 0.2--1.0 & UV/IR\tnote{k} \\ 
\bottomrule
\end{tabular}
\begin{tablenotes}
\item[a] Observation type used for SFRD estimate.
\item[b] Stacked SFHs from SED fits \citep[per][]{bellstedt20}.
\item[c] Cosmic IR background (CIB).
\item[d] {150~MHz radio LF.}
\item[e] Compiled by \citet{behroozi13,behroozi19}.
\item[f] Rescaled to Chabrier IMF by \citet{behroozi19}.
\item[g] UV LF.
\item[h] [O\,{\sc ii}] LF.
\item[i] 1.4~GHz radio LF.
\item[j] H$\alpha$ LF.
\item[k] Combined 2800~\AA\ and 24~$\upmu$m LFs.
\item[l] Combined $(15,24,70,150,870)~\upmu$m LFs.
\item[m] {3~GHz radio LF.}
\item[n] Estimated total IR LF.
\item[o] {IR SED modelling.}
\item[p] Rest-frame 24~$\upmu$m LF.
\end{tablenotes}
\end{threeparttable}
\end{table}

\subsection{Results}
\label{subsec:sfrd_result}

In Figure~\ref{fig:sfrd} we present our COSMOS2020-thresholded and model-thresholded estimates for the evolving cosmic SFRD, both computed in bins of width $\Delta z = 0.2$. These two estimates agree for $z \lesssim 1$, beyond which the model-thresholded estimate is $\sim\! 0.2$~dex higher than the COSMOS2020-thresholded one. We limit this analysis up to a redshift of $z=3.5$ in order to compare with the range best covered by the literature. 
All point estimates and functional representations of the SFRD in the figure have been converted to the Chabrier IMF \citep{chabrier03}. We present the set of literature point estimates overplotted in Table \ref{tab:sfrd_literature}, where we added recent data to the compilation assembled by \cite{behroozi13, behroozi19}.

We also show the canonical SFRD measurement derived by \cite{madau14} by fitting to a compilation of different estimates. We find that the qualitative shape and overall normalization ($0.08\pm0.01$ $\mathrm{M}_{\odot}\,\text{yr}^{-1}\,\text{Mpc}^{-3}$) of the \cite{madau14} result agrees well with our result. However, quantitatively there  is an important difference: we find that the peak of our SFRD at $z=1.3 \pm 0.1$ occurs $\Delta z \simeq 0.6$ later than their peak ($z\simeq1.9$). Using the fitting function from \citet{madau14} to represent the lower and upper envelopes of the \pc\ SFRD, we obtain
\begin{align}
\Psi_\text{lower}(z) &\simeq 0.003 \times \frac{(1+z)^{5.3}}{1 + [(1+z)/2.0]^{6.7}}~\mathrm{M}_{\odot}\,\text{yr}^{-1}\,\text{Mpc}^{-3},\label{eq:envelope-lower}\\
\Psi_\text{upper}(z) &\simeq 0.006 \times \frac{(1+z)^{4.7}}{1 + [(1+z)/2.0]^{6.1}}~\mathrm{M}_{\odot}\,\text{yr}^{-1}\,\text{Mpc}^{-3}.\label{eq:envelope-upper}
\end{align}
These expressions are a good representation of the \pc\ SFRD lower and upper envelopes for $0.2<z<3.5$. {The lower envelope in Equation \ref{eq:envelope-lower} corresponds to the lower edge of the COSMOS2020-thresholded curve in Figure \ref{fig:sfrd}, whilst the upper envelope in Equation \ref{eq:envelope-upper} corresponds to the upper edge of the model-thresholded curve. These consequently encompass the full extent of predictions from the \texttt{pop-cosmos} model.}

We further compare our results with another compilation study from \texttt{UniverseMachine} \citep{behroozi19}, an empirical framework which connects galaxy formation, and therefore SFHs, to dark matter halo assembly across cosmic time. Their empirical model is calibrated using a broad compilation of observed relationships for galaxy properties including SMFs, SFRs, and quenched fractions. Systematic offsets are added to these `true' model predictions to account for systematic observational and modelling uncertainties, resulting in a corrected `observed' model which aims to reproduce the observed relationships between galaxy properties out to $z \simeq 10$. Figure~\ref{fig:sfrd} shows that the \pc\ results lie between the `true' and `observed' models close to the SFRD peak at $z \simeq 1.3$, but agrees better with their `true' model at higher redshifts. {Many of the data points from the \citet{behroozi13, behroozi19} compilation lie above the \citet{madau14} and `true' \texttt{UniverseMachine} curves (as well as the \pc\ result), with these differences likely driven by the various methodological differences, uncorrected selection effects, and systematic errors present in the compiled observations (with the offset between \texttt{UniverseMachine}'s `observed' and `true' SFRD estimates aiming to represent these effects).}

We also show recent results from \cite{chiang25} who perform a comprehensive tomographic analysis of the cosmic infrared background (CIB). The CIB is a direct measure of star formation activity, as it probes the light re-emitted by dust originally emitted in the UV by young stars. The \cite{chiang25} results are derived from a $60\,\sigma$ detection of the evolving CIB spectrum over $0 < z < 4$, with minimal impact from cosmic variance. They compute an `IR' SFRD from the total co-moving IR luminosity in the CIB. They then compute a `total' SFRD which is obtained by adding integrals of UV galaxy luminosity functions compiled from the literature to the `IR' estimate. The `IR' estimate agrees with the \pc\ estimate, while their `total' estimate with UV LF corrections overshoot the \pc\ estimate as well as the \cite{madau14} curve. Separately, there is a jump of unknown origin in the \cite{chiang25} result between the $z\simeq1.4$ and $z\simeq1.6$ bins, which is present in both the `IR' and `total' estimates with the same magnitude, and therefore is unlikely to result from the UV LF corrections. The broader, later-peaking SFRD found by \pc\ appears to align qualitatively well with the \textit{Herschel} IR-derived SFRD shown in \citet{madau14} based on data from \citet{gruppioni13}, although these data were not included in the \citet{behroozi13, behroozi19} compilation.

Finally, we compare our result with a radio-continuum-based SFRD estimate derived from a large sample by \cite{leslie20}. The radio continuum provides a method of computing SFR values unimpacted by dust. \cite{leslie20} derive average SFRs by stacking 3~GHz radio continuum images of $\sim 200,000$ galaxies in the COSMOS field. Their SFRD estimates show a high level of agreement with our results, both in their normalization and the shape of the relation. In particular, their SFRD values at $z \simeq 1.3$ and $z \simeq 1.7$ are nearly equal within their uncertainties, consistent with the broad plateau of high star formation activity between $z \simeq 1$ and $z \simeq 2$ that we find. More recently, \cite{matthews24} have estimated the SFRD at $0.2 < z < 1.4$ using deep 1.4~GHz data. Their result, based on 3839 galaxies with high-confidence redshifts, shows a significantly different slope compared to \cite{madau14}, while being in agreement with our results and those of \cite{leslie20} where there is overlap in redshift. However, it is worth noting that other radio-based analyses report shallower slopes in this redshift range \citep{ocran20b,vandervlugt22}.

The later SFRD peak found by \pc\ could reflect better sensitivity to intermediate and low-mass star-forming galaxies at $z\simeq1$--$1.5$ through deep IR selection. UV-selected surveys may preferentially detect unobscured star formation in massive galaxies at $z\simeq2$, while missing dust-obscured star formation in lower-mass galaxies that becomes increasingly important at $z\simeq1$--$1.5$. The agreement with CIB measurements, which capture all dust-reprocessed starlight regardless of galaxy mass, {and the radio results of \cite{leslie20}, which are insensitive to dust,} supports the idea that a complete census of star formation may indeed peak later than previously thought.

\subsection{{Integrated SFRD}}
{As a further test of our SFRD, we use it to estimate a redshift evolving stellar mass density (SMD) by integrating the fitting function for our upper and lower SFRD bounds (Equations \ref{eq:envelope-lower} and \ref{eq:envelope-upper}). The time integral of the SFRD from a maximum redshift, $z_\text{max}$, to a lower redshift, $z$,  provides an estimate of the total amount of stellar mass formed up to that redshift, and thus the expected SMD, $\rho_*(z)$. This quantity can also be estimated by integrating a redshift-binned SMF across all masses, to estimate a total stellar mass at a particular $z$. When taking the former route (i.e.\ integrating the SFRD), the resulting SMD will count the total stellar mass formed. To obtain an estimate compatible with the latter route, a correction is needed to account for stellar mass loss (see e.g.\ \citealp{madau14}).}

\begin{figure}
    \centering
    \includegraphics[width=\linewidth]{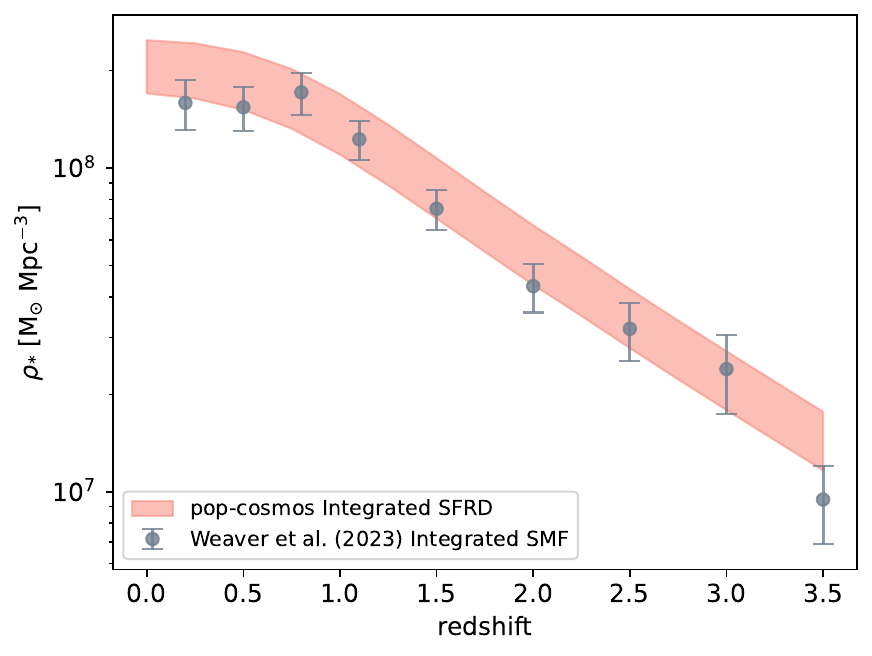}
    \caption{{The stellar mass density (SMD) for \texttt{pop-cosmos}, estimated by integrating the SFRD from Figure \ref{fig:sfrd} (and including a correction for mass loss based on our mock catalogue). Gray markers show the integrated SMF from \citet{weaver23}.}}
    \label{fig:smd}
\end{figure}

{In Figure \ref{fig:smd}, we show the \texttt{pop-cosmos} SMD, estimated via integrating our SFRD. We compare this result to the SMD presented by \citet{weaver23} for the COSMOS field, which they obtain from their SMFs in each of their redshift bins. We see good agreement, despite the two SMD estimates being obtained through very different analyses. To correct the \texttt{pop-cosmos} integrated SFRD for mass loss, we use our mock catalogue to estimate the median fraction of stellar mass remaining for galaxies at a given redshift\footnote{{Typically, a single `return fraction' is assumed at all redshifts, but this neglects the fact that galaxies' mass loss actually depends on redshift-evolving quantities such as metallicity (see \citealp{madau14}).}}. For each model galaxy, this is predicted by our SPS model. We find that the median fraction of mass remaining, $f_\text{remain}(z)$, is well modelled by a broken linear relation,}
\begin{equation}
    {
        f_\text{remain}(z)=\begin{cases}
            0.0415\,z + 0.5763 &\text{for }z<2.25\\
            0.0172\,z + 0.6347 &\text{for }z>2.25.
        \end{cases}
    }
\end{equation}
{We multiply this correction into our integrated SFRD to compute the SMD shown in Figure \ref{fig:smd}.The implied $z=0$ return fraction, $f_\text{return}(0)\equiv1-f_\text{remain}(0)=0.4237$, is in excellent agreement with the widely-used $f_\text{return}=0.41$ estimated by \citet{madau14} for a \citet{chabrier03} IMF.}

\section{Star-forming and Quiescent Subpopulations}
\label{sec:sf_q}

Understanding the transition from active star formation to quenching is a key aim of galaxy evolution studies \citep[for a review see, e.g.][]{man18, delucia25}. Separating observed galaxy populations into quiescent (Q) and star-forming (SF) subsamples is the starting point to investigating the evolutionary pathways and astrophysical processes, such as feedback-driven outflows \citep[see, e.g.][]{mcnamara07, fabian12} and/or other environmental processes \citep[see, e.g.][]{boselli22, alberts22} which cause this transition. The labelling of Q and SF subpopulations is commonly performed using colour--colour diagrams, or by using sSFR measures. A key advantage of our approach is the ability to compare the outcomes of these selection methods on an even footing. 

Here and for the remainder of this paper, we work with the COSMOS2020-thresholded mass-complete selection (Section \ref{subsec:data}) from the \pc\ mock catalogue.

\begin{figure*}
    \centering
    \includegraphics[width=0.9\linewidth]{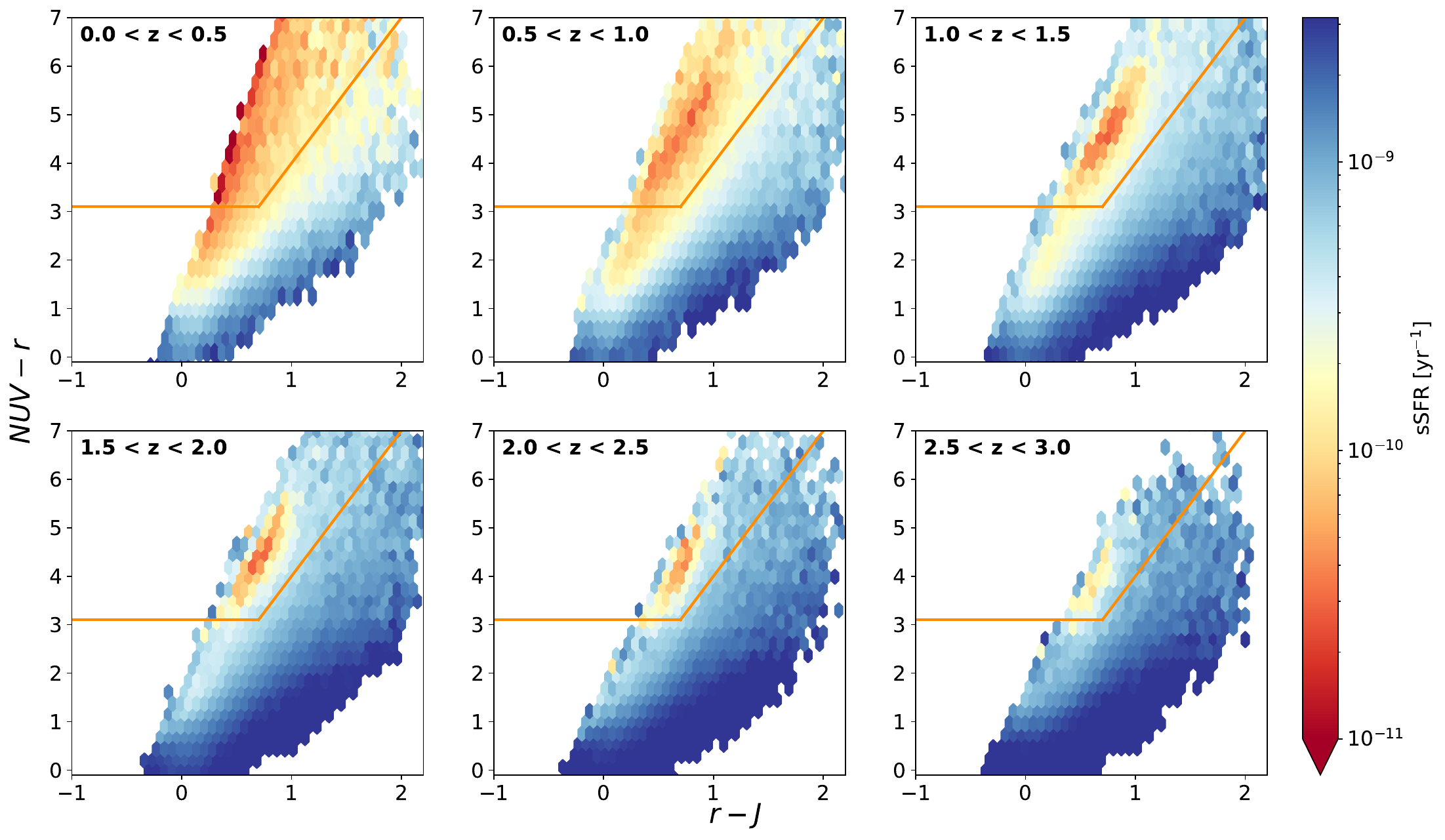}
    \caption{Rest-frame $NUVrJ$ colour--colour diagrams. Cells are shaded based on the median sSFR, and are only shaded when they contain $N>5$ galaxies. The orange line shows the SF/Q boundary from \protect\cite{weaver23}. In our colour-based analysis, Q galaxies are those to the upper left of the boundary. The sSFR range on the colourbar is limited to [$10^{-11}$, $10^{-8.5}$]~$\text{yr}^{-1}$.}
    \label{fig:nuvrj_ssfr}
\end{figure*}

\subsection{Labelling star-forming and quiescent galaxies}
\label{subsec:identify_sf_q}

The selection method for Q galaxies fundamentally affects our census of the quenched population. Colour-colour selection into Q and SF galaxies can be complicated by the difficulty of distinguishing between galaxies that are red due to old stellar populations as opposed to dust-reddened galaxies with young stellar populations. The SF/Q separation in colour-colour diagrams is commonly performed by using an empirically-defined selection boundary designed to minimize this contamination. A popular colour-colour combination is the $UVJ$ diagram \citep{wuyts07, williams09, tomczak14, leja19_uvj}, which uses the $U-V$ and $V-J$ colours. \cite{leja19_uvj} showed that the $UVJ$ diagram is prone to contamination from SF galaxies in the Q selection window, advocating the replacement of the $U$-band with shorter wavelength bands further in the UV, which correlate more strongly with sSFR. 

We now investigate the implications of selection based on the rest-frame $NUVrJ$ colour--colour diagram, and the canonical sSFR boundary for defining SF galaxies: $\text{sSFR}>10^{-11}\,\mathrm{yr}^{-1}$ \citep{ilbert10, ilbert13}. We note that redshift-dependent sSFR criteria have been introduced, rescaling the threshold based on the age of the Universe, $t_\text{Univ}(z)$. \citet{tacchella22a} suggest a Q selection using $\text{sSFR}<1/[20\times t_\text{Univ}(z)]$; this implies an sSFR cut of $10^{-11}\,\text{yr}^{-1}$ at $z\simeq1.3$, and a factor of $\sim2\times$ lower and higher at $z\simeq0.5$ and $z\simeq3.0$, respectively. We thus expect that redshift-dependent selection would not significantly change our conclusions.

\begin{figure*}
    \centering
    \includegraphics[width=0.995\linewidth]{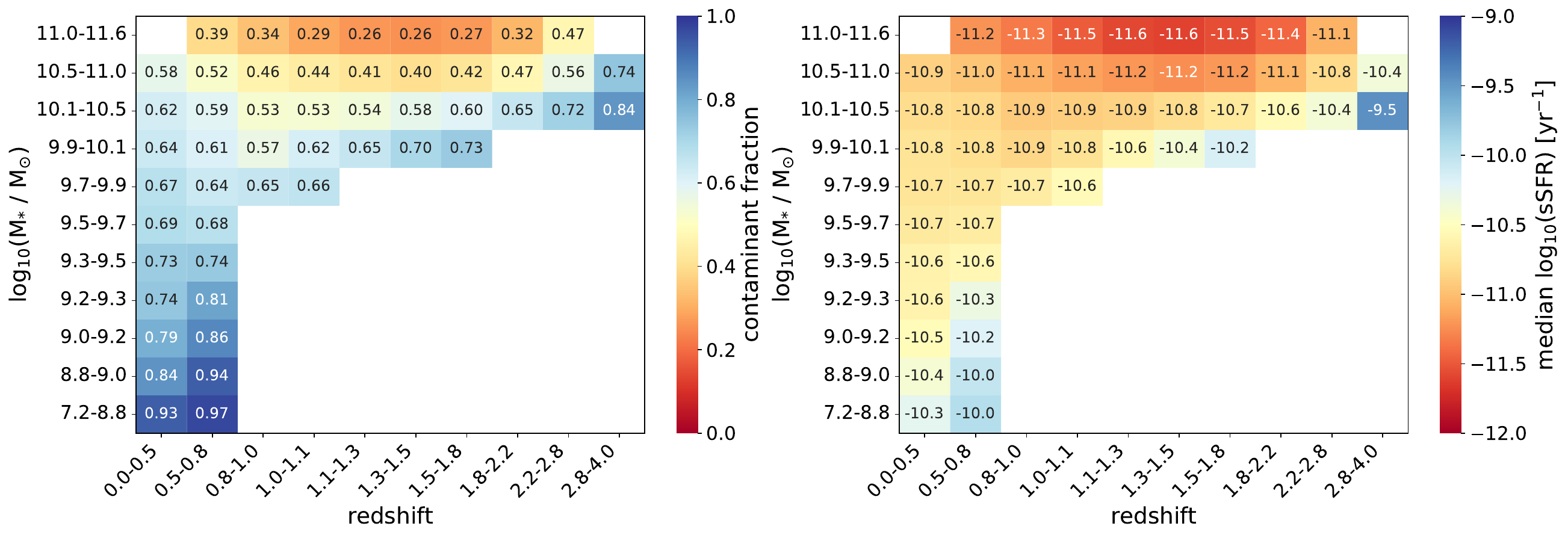}
    \caption{\textbf{Left:} Contaminant fraction -- i.e.\ false discovery rate, $\mathrm{FP}/(\mathrm{TP}+\mathrm{FP})$ -- for Q galaxies identified via $NUVrJ$, in bins of redshift and stellar mass above our mass-completeness limit. Cells above the mass-completeness threshold are populated if they contain at least 1000 galaxies. \textbf{Right:} Same as left panel, but with bins shaded based on the median log$_{10}$(sSFR) of the galaxies in the bin. Each redshift (column) and stellar mass (row) bin contains 10 percent of the sample, except for the two most massive bins which contain 10 percent between them. }
    \label{fig:f_contam}
\end{figure*}

We compute rest-frame $NUVrJ$ photometry for the \pc\ galaxies as described in Section \ref{subsec:restframe} and \ref{subsec:data}. Throughout the paper we use the colour--colour boundary used by \citet{weaver23} to label galaxies as `$NUVrJ$-selected' SF/Q. Figure \ref{fig:nuvrj_ssfr} shows the  median sSFR of galaxies in different parts of colour--colour space, with the \citet{weaver23} selection boundary overlaid. Our analysis reveals that $NUVrJ$-selected quiescent samples contain galaxies with sSFR up to $10^{-9}$~yr$^{-1}$, two orders of magnitude above the canonical quenching threshold of $\text{sSFR}=10^{-11}$~yr$^{-1}$. 

In the left panel of Figure \ref{fig:f_contam} we show the contaminant fraction in the form of the false discovery rate $\mathrm{FP}/(\mathrm{TP}+\mathrm{FP})$ in mass and redshift bins, where a contaminant (false positive) is defined as a galaxy which is classified as SF by the sSFR selection but as Q by the $NUVrJ$ selection. Contamination is lowest at the high mass end of the distribution but even then, we see that the minimum contamination is $\sim26$ percent false positives. In the right panel of Figure \ref{fig:f_contam} we also show the median sSFR in the same bins. It is worth bearing in mind that our sSFR selection criterion is for very high-confidence Q galaxies; relaxing this to $\text{sSFR}< 10^{-10.5}$ yr$^{-1}$ would yield lower contaminant fractions down to $10^{-10}$ yr$^{-1}$, at the expense of including galaxies transitioning to quiescence. 

Upon examining the false positives that are flagged as Q by the $NUVrJ$ criterion while having $\text{sSFR}> 10^{-11}$~yr$^{-1}$, we find that their dust content (parametrized in \pc\ by the diffuse dust attenuation optical depth, $\tau_2$) is strongly correlated with sSFR.  This is seen in \pc\ for the general galaxy population \citep[see][]{alsing24, thorp25b} and is a well-known relation in other observational and theoretical studies \citep[e.g.][]{garn10, chevallard13, zahid13, sommovigo25}. These galaxies enter the $NUVrJ$ selection boundary most prominently at $NUV-r\gtrsim5$ and $r-J\gtrsim1$ for redshifts $z\gtrsim0.5$. Modification of the selection boundary in this region to exclude this region, or adoption of a redshift dependent boundary, may lead to a better consistency with sSFR-based selection.

\begin{figure*}
    \centering
    \includegraphics[width=\linewidth]{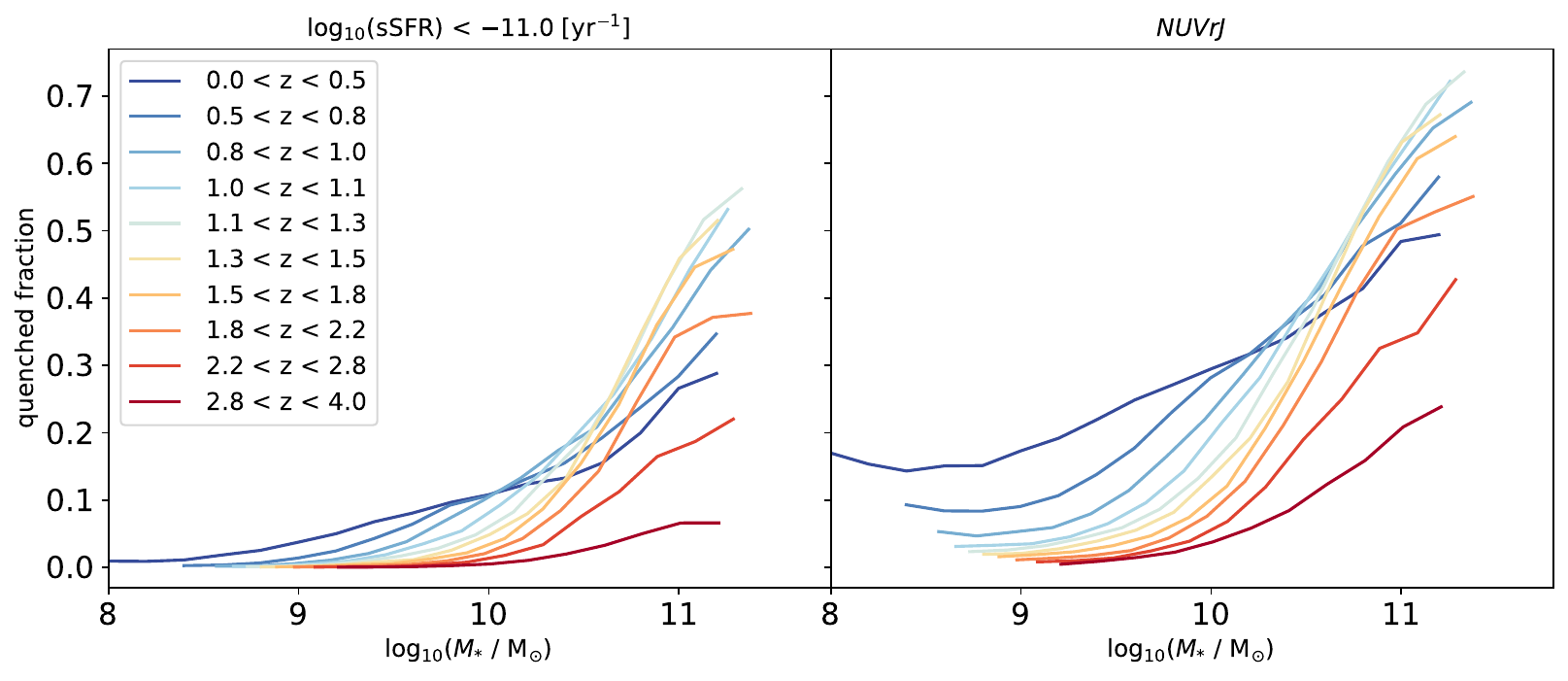}

    \caption{Quenched fraction as a function of stellar mass in bins of redshift. Each redshift bin contains 10 percent of the sample. \textbf{Left:} SF/Q selection based on an sSFR threshold of $10^{-11}\,\text{yr}^{-1}$. \textbf{Right:} SF/Q selection based on $NUVrJ$ diagram.}
    \label{fig:f_q_zbins}
\end{figure*}

\begin{figure*}
    \centering
    \includegraphics[width=\linewidth]{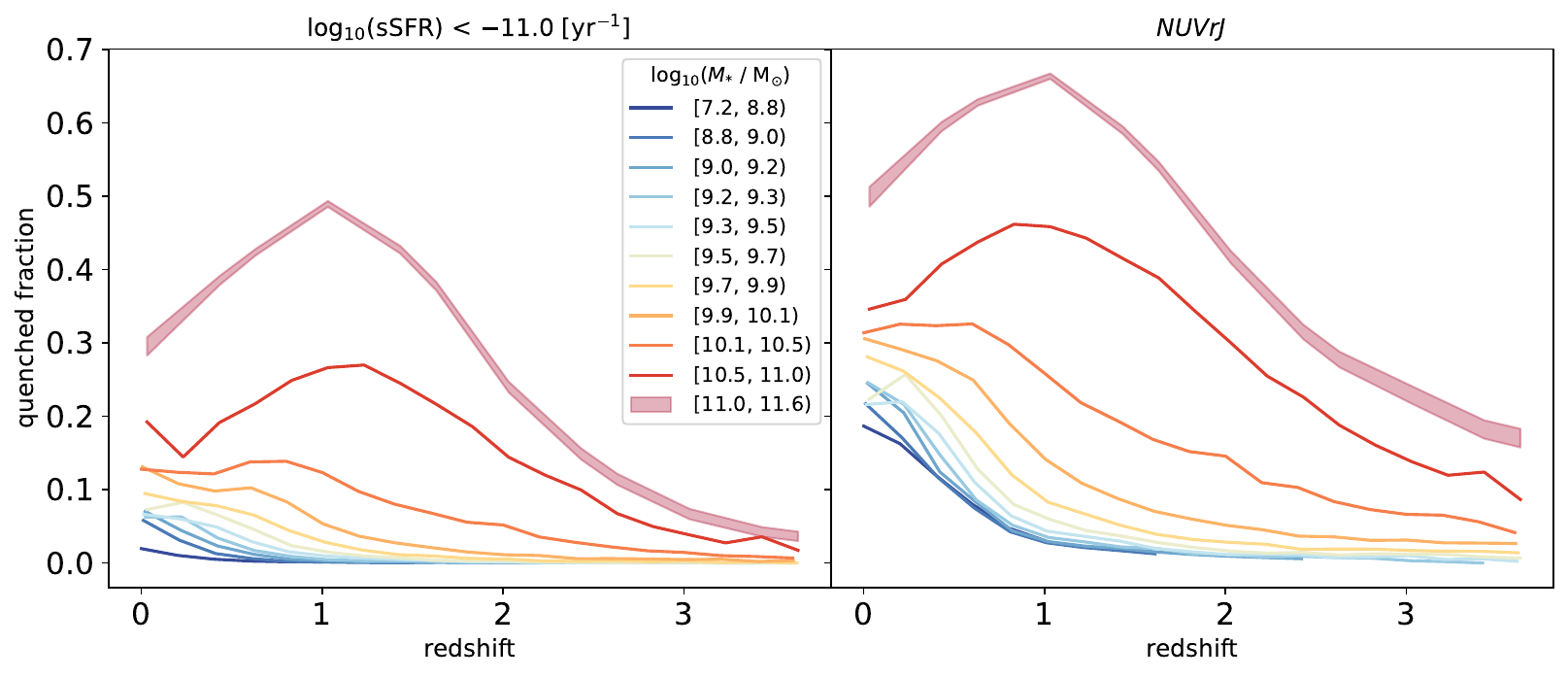}
    \caption{The quenched fraction as a function of redshift in bins of stellar mass. Each stellar mass bin contains 10 percent of the sample, except for the two most-massive bins, which contain 10 percent between them. \textbf{Left:} SF/Q selection based on an sSFR threshold of $10^{-11}\,\text{yr}^{-1}$. \textbf{Right:} SF/Q selection based on $NUVrJ$ diagram. Estimated cosmic variance for a COSMOS-sized field is shown as a shaded region for the most massive bin.}
    \label{fig:f_q_mbins}
\end{figure*}

\subsection{Quenched fractions}
\label{subsec:qf}

A key derived quantity that can be immediately computed following the SF/Q classification is the quenched fraction \citep[e.g.][]{baldry06, peng10, alberts22}, which we visualize as a function of stellar mass and redshift in Figures \ref{fig:f_q_zbins} and \ref{fig:f_q_mbins}. The contamination discussed in Section \ref{subsec:identify_sf_q} propagates directly into this derived quantity. The overall normalization of the quenched fraction curves is higher for the $NUVrJ$: for instance at $z\simeq 1$, the quenched fraction reaches $\sim40$ percent at $10^{10.5}$ M$_{\odot}$ for the $NUVrJ$ selection, compared with $\sim 25$ percent for the sSFR selection. For stellar masses $10^{10.5}$--$10^{11}~\mathrm{M}_\odot$, the quenched fraction increases from $\sim 25$~percent at $z =2.5$ to $\sim45$~percent by $z = 1.0$ for the $NUVrJ$ selection, while the sSFR selection predicts an increase from $\sim10$~percent to $\sim25$~percent over the same redshift range. We show on Figure~\ref{fig:f_q_mbins} the estimated cosmic variance in the most massive stellar mass bin, computed as in Section \ref{subsec:sfrd_unc}. Whilst we would expect this to be the bin where this effect is strongest, at all $z\lesssim3$ the cosmic variance is insignificant relative to the trends seen in the figures.

\begin{figure*}
    \centering
    \includegraphics[width=\linewidth]{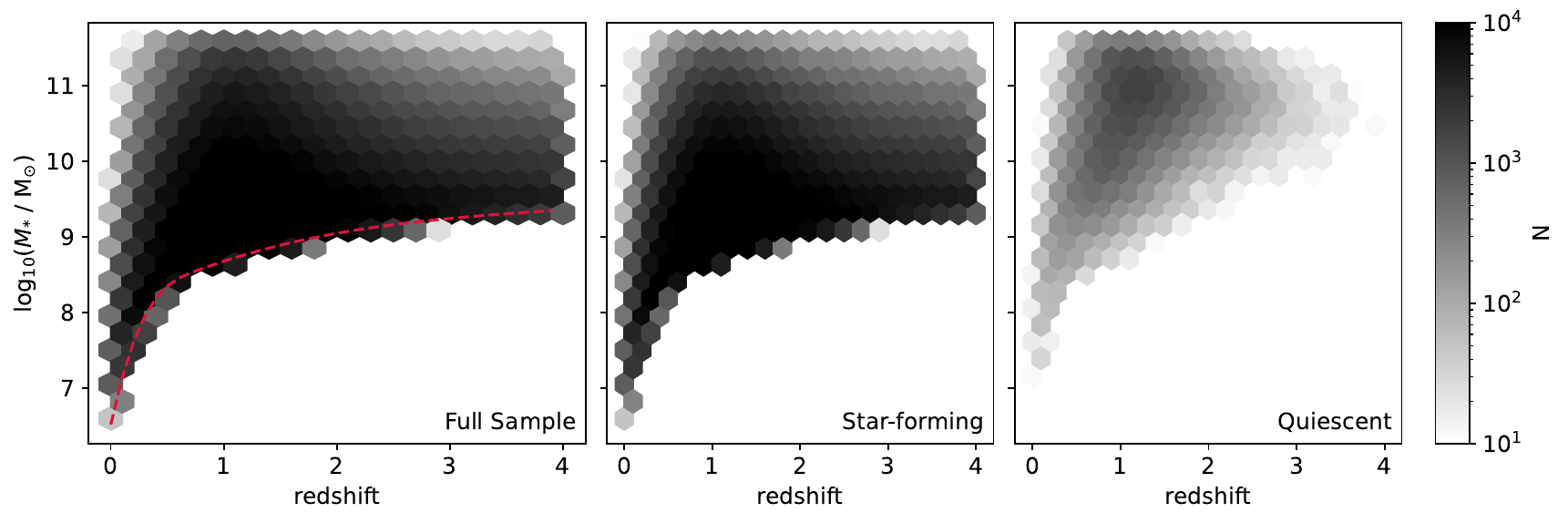}
    \caption{The stellar mass distribution of the full (left), star-forming (middle), and quiescent galaxies from \pc. Cells are shaded by number of galaxies. Only galaxies above our mass completeness threshold (red curve, left panel) are included.}
    \label{fig:mstar_z}
\end{figure*}

In Figure~\ref{fig:f_q_zbins} we show the quenched fraction as a function of stellar mass and in bins of redshift for the $NUVrJ$ and sSFR selections. The redshift bins are chosen to each contain 10 percent of the sample. The overall trends seen in both selections are similar: the quenched fraction increases with mass and decreases with redshift. Most clearly in the sSFR-based SF/Q selection, we see that for galaxies less massive than $\sim10^{9.5}\,\mathrm{M}_\odot$, the quenched fraction is very low ($\sim10$ percent or less) at all redshifts. Conversely, for galaxies more massive than $\sim10^{10.5}\,\mathrm{M}_\odot$, the quenched fraction is $\sim20$ percent or higher for all $z\lesssim2$. The mass range between $\sim10^{9.5}$ and $10^{10.5}\,\mathrm{M}_\odot$ thus appears as an important transitional range, implying that the quenching mechanism(s) at play are neither a sharp threshold-like phenomenon (e.g.\ major mergers; \citealp{mihos96, springel05, dimatteo05, hopkins08a, hopkins08b, johansson09, faisst17, ellison24, heckman24}), nor extremely gradual (e.g.\ starvation of gas; \citealp{bekki02, vandenbosch08, peng15}). Mergers as a primary quenching channel have also been disfavoured by a number of recent theoretical \citep[e.g.][]{weinberger18, rodriguez19, quai21, quai23} and observational \citep[e.g.][]{weigel17, ellison18, inoue24} works (although see discussion in \citealp{zheng22}, and observational evidence of rapid quenching from e.g.\ \citealp{socolovsky18, carnall18, belli19, wild20, forrest20, tacchella22a, park23_rapid, park24_rapid}).

Figure \ref{fig:f_q_mbins} reveals the redshift evolution of the quenched fraction in bins of stellar mass, with each bin containing 10 percent of the sample (apart from the two most massive bins, which contain 10 percent between them). Although there is  again a systematic difference in the normalization of the curves between the $NUVrJ$-based and sSFR-based definitions of the SF/Q split, the qualitative behaviour is very similar. For both selections, above stellar masses of $10^{10}\,\mathrm{M}_\odot$ there is a factor of $\sim2$ growth in the quenched fraction {between $z\simeq2$ and $z\simeq1$}. In the sSFR-based selection, which is less diluted by contaminants, this effect is particularly strong. For stellar mass bins below $10^{10}\,\mathrm{M}_\odot$, the quiescent fractions are much more tightly grouped at a given redshift. In the sSFR-based selection, the quiescent fractions for stellar masses $\lesssim10^{10}$ are  below $\sim15$~percent at all redshifts. This aligns well with a picture of galaxy evolution where star formation is increasingly suppressed above stellar masses $\sim10^{10.5}\,\mathrm{M}_\odot$ due to AGN activity \citep[see, e.g.][]{bower17}. We shall revisit this connection in Section \ref{sec:discussion}. 

{We notice that the quenched fraction in Figure \ref{fig:f_q_mbins} tends to decline between $z\simeq1$ and $z\simeq0$, for both the sSFR- and $NUVrJ$-selected samples. This is only prominent in the highest stellar mass bins $\log_{10}(M/\mathrm{M}_\odot)\geq10.5$. This feature has been present and stable in both calibrations of the \pc\ model. This result cannot simply be a statistical fluctuation; while the calibration dataset from COSMOS2020 has its lowest volume at low redshift, the cosmic variance is still relatively small (as seen in Figure \ref{fig:f_q_mbins}). One possibility is that the low-$z$ volume sampled in this relatively small area survey is not representative of the galaxy population that would be seen in wide-area surveys, particularly the relatively rare population of massive quiescent galaxies. We therefore expect this to be the regime where the calibration is most uncertain. As noted in \citet{thorp25b}, future calibrations of the model that combine deep COSMOS data \citep[from][]{shuntov25} with a wide-area survey -- e.g.\ the Kilo Degree Survey \citep{wright24}, HSC \citep{aihara22}, the DESI Bright Galaxy Survey \citep{hahn23_bgs, abdul25}, or the Galaxy and Mass Assembly survey \citep{driver22} -- will likely improve constraints in this regime. On the other hand, if we take this result at face value, an exciting explanation is the rejuvenation of star formation activity especially for massive galaxies at $z<1$. We leave the investigation of these possibilities to future work.}

Figures \ref{fig:f_q_zbins} and \ref{fig:f_q_mbins} show that more massive galaxies quench earlier, and that quenching efficiency indicates a somewhat gradual (as opposed to a threshold-like) physical process. These trends agree with previous estimates of the quenching fraction from observations \citep[e.g.][]{muzzin13, bauer13, ilbert13, weaver23}, empirical models \citep[e.g.][]{moster18, behroozi19}, and cosmological simulations \citep[e.g.][]{chaikin25}. Examining our sSFR-based quenched fractions at $z\simeq1$, our estimate of $\sim50$ percent quenched in the highest mass bin aligns very well with the $NUVrJ$-based observational estimates from \citet{weaver23}, and with the recent simulations from \citet{chaikin25}. However, for the lower stellar mass bins, our sSFR-based quenched fraction estimates tend to be lower than \citet{weaver23} by a factor of up to $\sim2\times$; e.g.\ for $10^{9.5}$--$10^{10}$, $10^{10}$--$10^{10.5}$, and $10^{10.5}$--$10^{11}\,\mathrm{M}_\odot$, they estimate quenched fractions of $\sim10$, 30, and 40 percent, respectively. In these lower-mass bins, our $NUVrJ$-based calculations align more closely with the results of \citet{weaver23}, as expected given the same selection methodology. We estimate that using an $NUVrJ$ selection yields up to $\sim 20$ percent over-estimates in the quenched fractions compared with sSFR selection across the $0<z<3.5$ in the mass range we consider. 

Overall, our results point to a picture of more ongoing star formation to later cosmic times, and less efficient quenching, than conclusions typically reached in earlier literature \citep[e.g.][]{muzzin13}. However, at $z\simeq1$ our results for $\sim10^{11}\,\mathrm{M}_\odot$ galaxies align well with the $\sim50$ percent quenching fraction estimated by e.g.\ \citet{moster18}, \citet{weaver23} and \citet{chaikin25}, and with the $\sim10$~percent quenching of $\sim10^{10}\,\mathrm{M}_\odot$ galaxies estimated by \citet{moster18} and \citet{chaikin25} at the same redshift.

\begin{figure*}
    \centering
    \includegraphics[width=\linewidth]{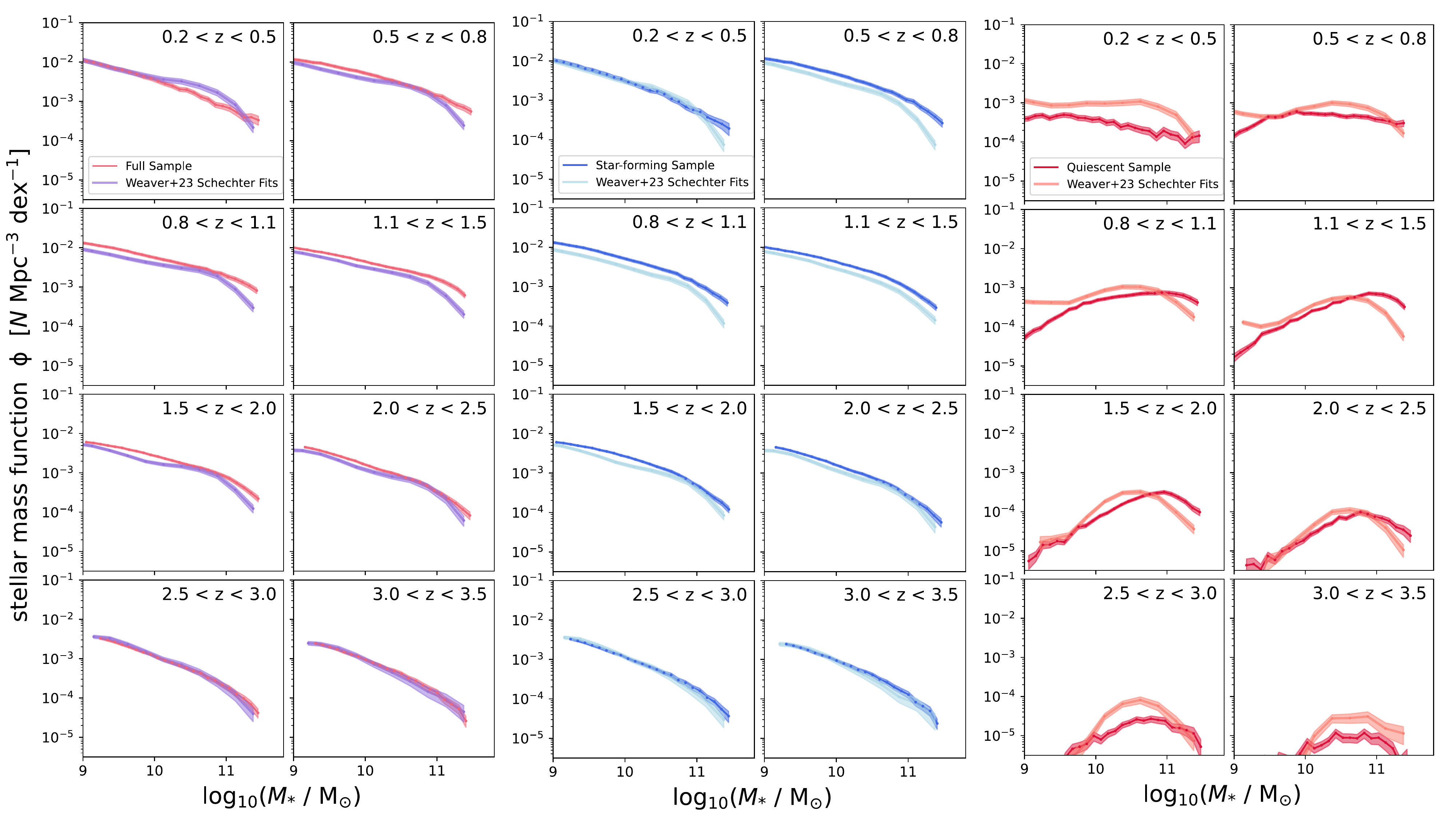}
    \caption{The SMF of the full (left), star-forming (middle), and the quiescent galaxies (right) in bins of redshift. Shaded regions show uncertainty due to cosmic variance and Poisson noise. We overplot the Schechter functions from \citet{weaver23}, including their reported uncertainty due to Poisson noise, cosmic variance, and stellar mass uncertainty from SED fitting. Redshift bins are from \citet{weaver23}. Only galaxies above our mass-completeness threshold (see Section \ref{subsec:data}) are included.}
    \label{fig:smf_separate}
\end{figure*}

\section{Stellar Mass Properties of Star-forming and Quiescent Populations}
\label{sec:smf}

The SMF provides a comprehensive view of the stellar mass assembly of galaxy populations \citep{muzzin13, ilbert13, weigel16, davidzon17, weaver23, shuntov25}. Having established the impact of using different SF/Q selection methods as a function of mass and redshift, we now investigate the SMFs of Q and SF subpopulations in \pc\ based on the sSFR criterion, comparing with results from \cite{weaver23}, which are based on $NUVrJ$ colour selection and also obtained from the COSMOS2020 catalogue. 

We begin by plotting the stellar mass distribution as a function redshift in Figure~\ref{fig:mstar_z} for the full, star-forming, and quiescent samples. Only galaxies above our mass-completeness threshold (see Section \ref{subsec:data}) are included. Quiescent galaxies first appear at $z \simeq 3.5$, exclusively at high stellar masses $> 10^{10.5}$\,M$_\odot$. The Q population density peaks at the mass scale $10^{10.5}$ M$_\odot$ and $z \simeq 1.0$. Below this redshift, the high-mass SF subpopulation declines while the Q population continues to grow, indicating ongoing quenching of massive galaxies.

\begin{figure*}
    \centering
    \includegraphics[width=\linewidth]{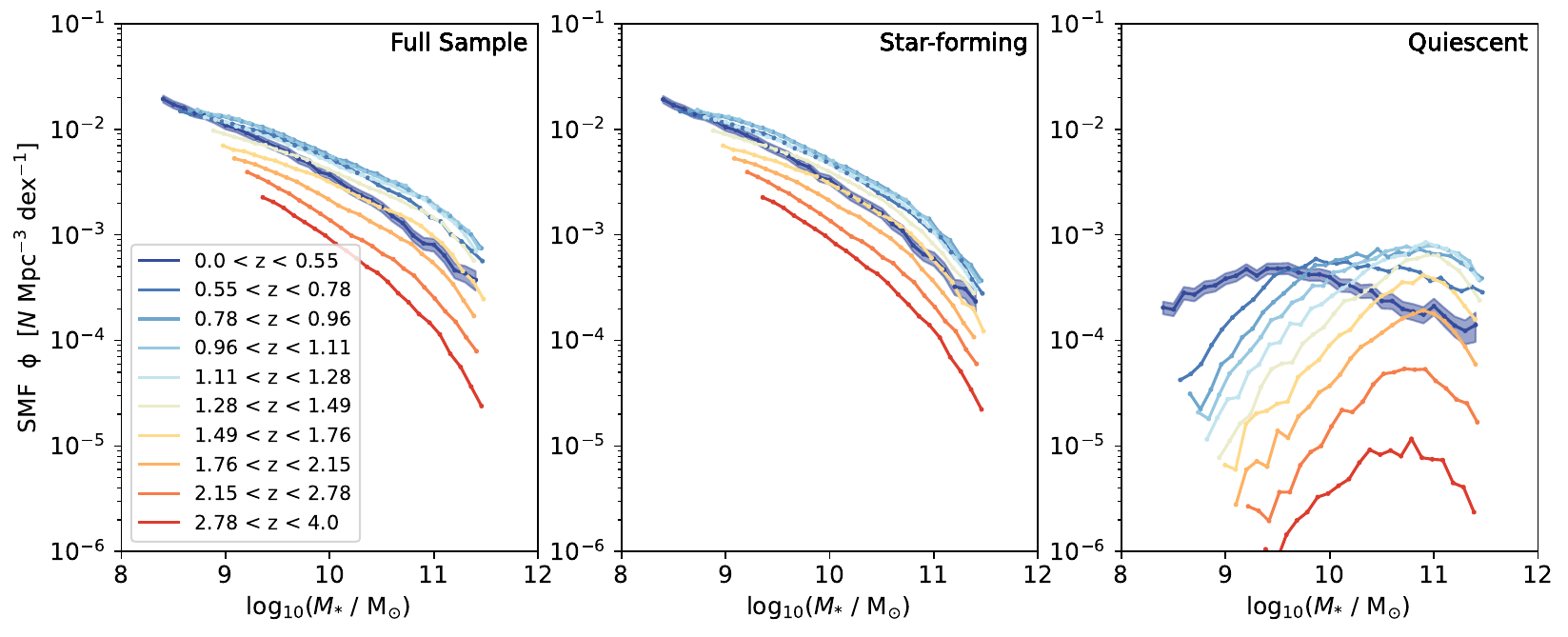}
    \caption{The redshift-binned SMF of the full (left), star-forming (middle), and quiescent (right) samples. Redshift bins contain 10 percent of the full sample. Shaded regions show uncertainty due to cosmic variance and Poisson noise. Only galaxies above our mass-completeness threshold (see Section \ref{subsec:data}) are included.}
    \label{fig:smf_joint}
\end{figure*}

Figure \ref{fig:smf_separate} shows SMFs in bins of redshift for the full, SF, and Q galaxies. We use the same redshift bins as \cite{weaver23} and overplot their results, which were obtained using Schechter function fits to stellar mass estimates based on \texttt{LePhare} \citep{arnouts99, ilbert06, ilbert09}. As previously noted, the SF/Q classification from \cite{weaver23} made use of $NUVrJ$ colour-based selection, in contrast with our sSFR-based selection for the \pc\ results displayed here. \cite{weaver23} evaluate their uncertainty as a quadrature sum of contributions from Poisson noise, cosmic variance, and uncertainty due to the estimation of stellar mass from SED fitting. The \pc\ uncertainties are obtained following the procedure described in Section \ref{subsec:sfrd_unc} as the quadrature sum of the cosmic variance and Poisson noise.

The SMFs in Figure \ref{fig:smf_separate} quantify how SF/Q selection affects our understanding of galaxy populations. For the full sample, our SMFs agree with \cite{weaver23} to within $\sim0.2$~dex across $0.2<z<3.5$ for all galaxies $\lesssim10^{11}\,\mathrm{M}_\odot$, and to within $\sim0.5$~dex for the tail of the SMF at $\gtrsim10^{11}\,\mathrm{M}_\odot$. This agreement validates the \pc\ normalization described in Section \ref{subsec:norm}.

The critical difference emerges in the Q sample at low masses. With sSFR selection, the Q galaxy SMF shows a power-law decline at mass scales $<10^{9.5}$\,M$_\odot$ with logarithmic slope $\alpha=1.1$. In contrast, the $NUVrJ$-selected SMF from \citet{weaver23} shows an upturn with $\alpha=-0.5$. This sign change has significant implications: at $10^{9}$ M$_\odot$ and $z \simeq 0.5$ $NUVrJ$ selection yields a Q galaxy number density of $6\times 10^{-4}~\text{Mpc}^{-3}\,\text{dex}^{-1}$ while sSFR selection gives $2\times 10^{-4}~\text{Mpc}^{-3}\,\text{dex}^{-1}$, a factor of $\sim\!3$ difference. This discrepancy directly can be traced back to the contamination identified in Section~\ref{subsec:identify_sf_q}. The up to 20 percent of $NUVrJ$-selected Q galaxies with sSFR $> 10^{-11}$ yr$^{-1}$ (Section~\ref{subsec:identify_sf_q}) artificially inflates the low-mass Q population\footnote{{If we apply an $NUVrJ$ selection to the \pc\ model draws, we are able to reproduce the upturn in the low-mass Q galaxy SMF that is seen in the COSMOS results from \citet{weaver23}.}}. These are dusty SF galaxies misclassified due to reddening, not genuinely quenched systems. Recent spectroscopic analysis \citep{mintz25} has identified a deficit in low-mass Q galaxies when using sSFR-based classification, which is consistent with our results.

There are other differences that may be due to environmental effects specific to the COSMOS field: as noted in \cite{thorp25b}, there are known large-scale structures at $z\simeq0.35$ \citep{scoville07_lss, sochting12, cherouvrier25} which introduce an overabundance of massive Q galaxies around $10^{10}$--$10^{11}\,\mathrm{M}_\odot$. Due to the way \pc\ is calibrated on distributions of colours and fluxes, it is less sensitive to such line-of-sight effects when used as a generative model \citep{thorp25b}. 

We now turn to the evolutionary trends in the SMF. Figure \ref{fig:smf_joint} shows the SMF evolution of the full, SF and Q populations, with each redshift bin containing 10 percent of the sample. We see three key trends. First, the total SMF normalization increases by a factor of 6 from $z\simeq 3.4$ to $z\simeq 0.7$ driven primarily by star formation at stellar mass $< 10^{10}$\,M$_\odot$. The second notable trend is the rapid assembly of the Q population. Between $z\simeq 3.4$ and $z\simeq 0.7$ the Q galaxy number density at mass scale $10^{10.5}$ M$_\odot$ increases from $9\times 10^{-6}$ to $4\times 10^{-4}~\text{Mpc}^{-3}\,\text{dex}^{-1}$, a factor of 50. In this period, the SF population at the same mass scale increases by a factor of 5, confirming that quenching, not just mass growth, drives the Q population assembly. Finally, we observe that the low-mass slope of the Q SMF flattens from $\alpha=1.1$ at $z\simeq3.4$ to $\alpha=0.4$ at $z\simeq 0.2$. This flattening indicates that the quenching efficiency decreases towards lower masses over cosmic time, consistent with environmental quenching becoming more important at late times when group/cluster environments are more common \citep[see e.g.][]{muzzin12, balogh16, matharu21, alberts22}. We observe a minor increase of $0.5$ dex in the peak mass scale of the Q SMF as it evolves from $z \simeq 3$ to $z\simeq 1$, which could be explained by gas-poor minor mergers which do not trigger star formation \citep[e.g.][]{naab09, bezanson09, newman12, zahidgeller17, seuss23}. 

{The SMF in the lowest redshift bin for Q galaxies is dissimilar to the higher redshift bins in Figure \ref{fig:smf_joint}. We postulate that this behaviour at $z<0.55$ shares a common origin with the downturn in the high-mass quiescent fraction at low-$z$ that is seen in Figure \ref{fig:f_q_mbins} and discussed in Section \ref{sec:sf_q}. As mentioned there, a targeted low-$z$ calibration of \pc\ with a wide-area dataset will be necessary to check if this trend is arising from the specific massive Q population seen in this small area of sky at low-$z$. An alternative explanation is the rejuvenation of massive galaxies at low redshift. We will investigate this in more detail in future work.}

The mass integral of the Q (or SF) population growth rate above the typical quenching mass scale $\sim 10^{10.5}$ M$_\odot$ \citep{bower17} can be roughly estimated by differencing the SMFs in redshift slices. We find that the stellar mass \emph{quenching} rate around the SFRD peak is roughly constant between $z\simeq1.4$ and $z\simeq1.2$ at $\sim 0.05~\mathrm{M}_\odot\,\text{yr}^{-1}\,\text{Mpc}^{-3}$, while over the same redshift interval the stellar mass \emph{formation} rate decreases from $\sim 0.04$ to $\sim 0.03~\mathrm{M}_\odot\,\text{yr}^{-1}\,\text{Mpc}^{-3}$. This implies that  quenching drives the decline in cosmic star formation since $z\simeq 1$ seen in Figure \ref{fig:sfrd}. We now investigate the SFH of the SF and Q populations at the peak of our SFRD in more detail. 

\begin{figure*}
    \centering
    \includegraphics[width=0.95\linewidth]{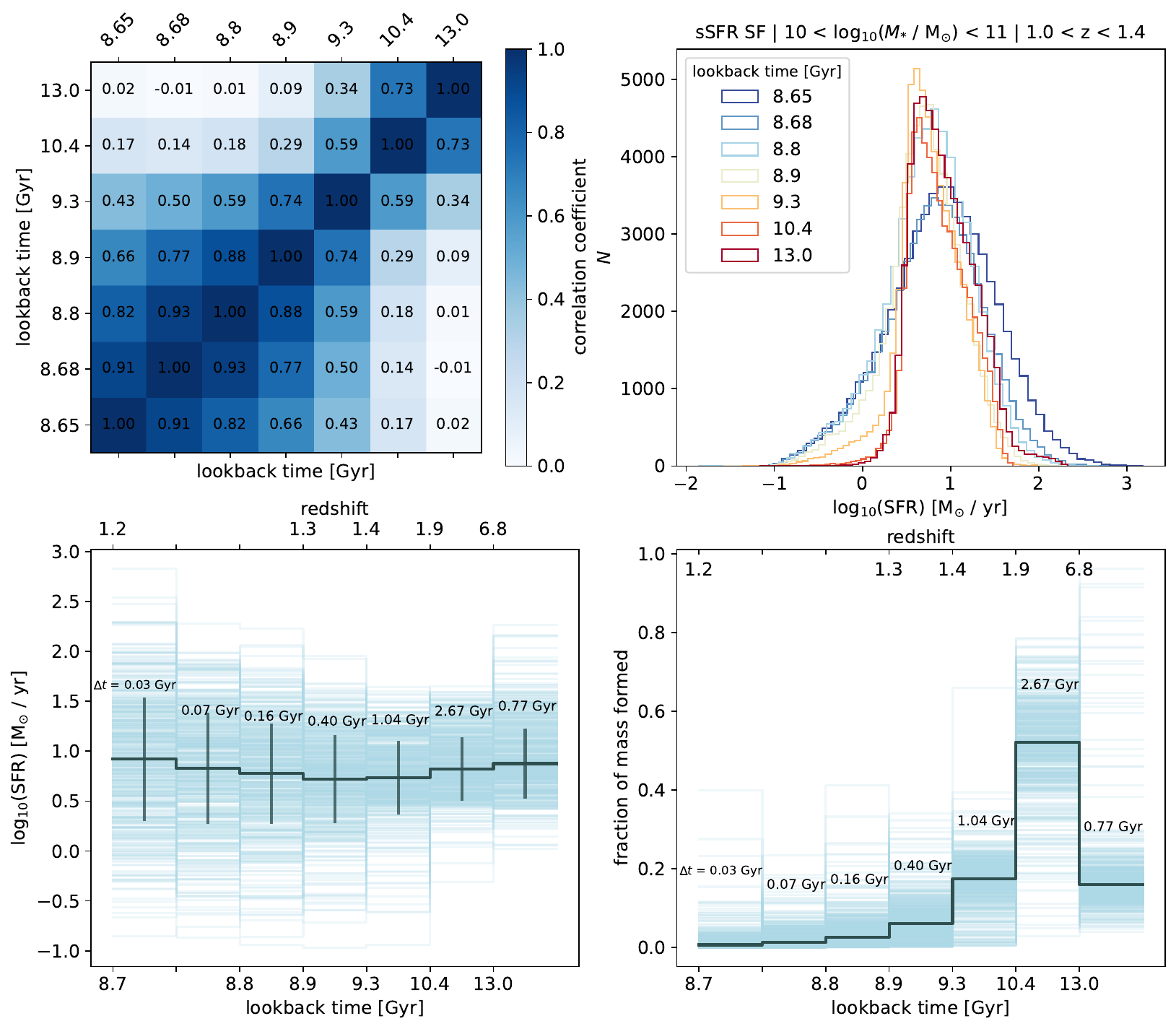}
    \caption{Star formation and stellar mass assembly history of the SF sample at $1.0<z<1.4$, with $10<\log_{10}(M_*/\mathrm{M}_\odot)<11$ and $\log_{10}(\text{sSFR}/\text{yr}^{-1})>-11$. \textbf{Top left:} Correlation matrix between the SFH bins of this sample. Colourbar range is fixed to [0, 1]. \textbf{Top right:} The SFR distributions in the 7 SFH bins. \textbf{Bottom left:} Median SFR and standard deviation per SFH bin for this sample (black), and representative SFHs for 500 randomly-selected individual galaxies (blue). Bins are plotted with equal width for visual clarity; their lengths in Gyr are annotated above the markers. \textbf{Bottom right:} Fraction of total stellar mass formed in each SFH bin. Lookback times quoted on the panels are the leftmost edges of the SFH bins, evaluated for the median redshift of the sample.}
    \label{fig:sf_massive}
\end{figure*}

\section{Star Formation Histories}
\label{sec:sfh}

Constraining the SFH of galaxy populations from individual SED fits to photometric (or even spectroscopic) data is extremely challenging, as observations are more sensitive to recent star formation than to older stellar populations. Population-level `hierarchical' models such as \pc\ can offer a unique window into the distribution of galaxy SFHs, by leveraging the partial pooling of information that is achieved when simultaneously modelling a large, representative sample. In this way, the problem of constraining individual galaxies' SFHs is sidestepped, and one directly learns the distribution of plausible SFHs in the whole galaxy population (for further discussion of these issues, see \citealp{wang25}). {The investigation here into SFHs is somewhat orthogonal to the population-level constraints on the star-forming main sequence presented in \citet{alsing24} and \citet{thorp25b}. Where those analyses looked at the instantaneous (i.e., averaged over the most recent 100~Myr) SFR of model galaxies `observed' at all redshifts, here we take the opposite approach of looking at the full SFHs of model galaxies `observed' in a single narrow redshift slice.}

{In such a study, one might be concerned about the information loss in the `observed' SEDs about star formation at progressively earlier epochs, due to observational effects such as outshining as well as SPS parameter degeneracies. 
However, this limitation can be avoided by selecting SF and Q subpopulations in the same redshift and mass slices, and \emph{comparing} their SFH. In this case, since both subpopulations are subject to identical information loss due to the above effects, any differences in the properties of the SFH subpopulations observed must arise due to the astrophysics of galaxy evolution. In this section we use this strategy to compare subpopulations of Q and SF galaxies selected in mass slices at the redshift where the \pc\ SFRD peaks.}

\subsection{The SFH of massive SF/Q galaxies from \pc}
\label{subsec:sfh_sf_q}

In this section we investigate the star formation histories of select galaxy populations from \pc. For this study we select subsamples from the \pc\ mock catalogue (Section~\ref{subsec:data}) within narrow stellar mass and redshift windows, classifying the galaxies in these subsamples as SF/Q using the sSFR-based criterion discussed in Section \ref{sec:sf_q}. For the galaxies in each subsample, we examine the SFHs, represented as seven bins of SFR spanning their lifetimes (see Section \ref{subsec:model} for the details of the \pc\ non-parametric SFH, which is based on \citealp{leja19_sfh}). 

\begin{figure*}
    \centering
    \includegraphics[width=\linewidth]{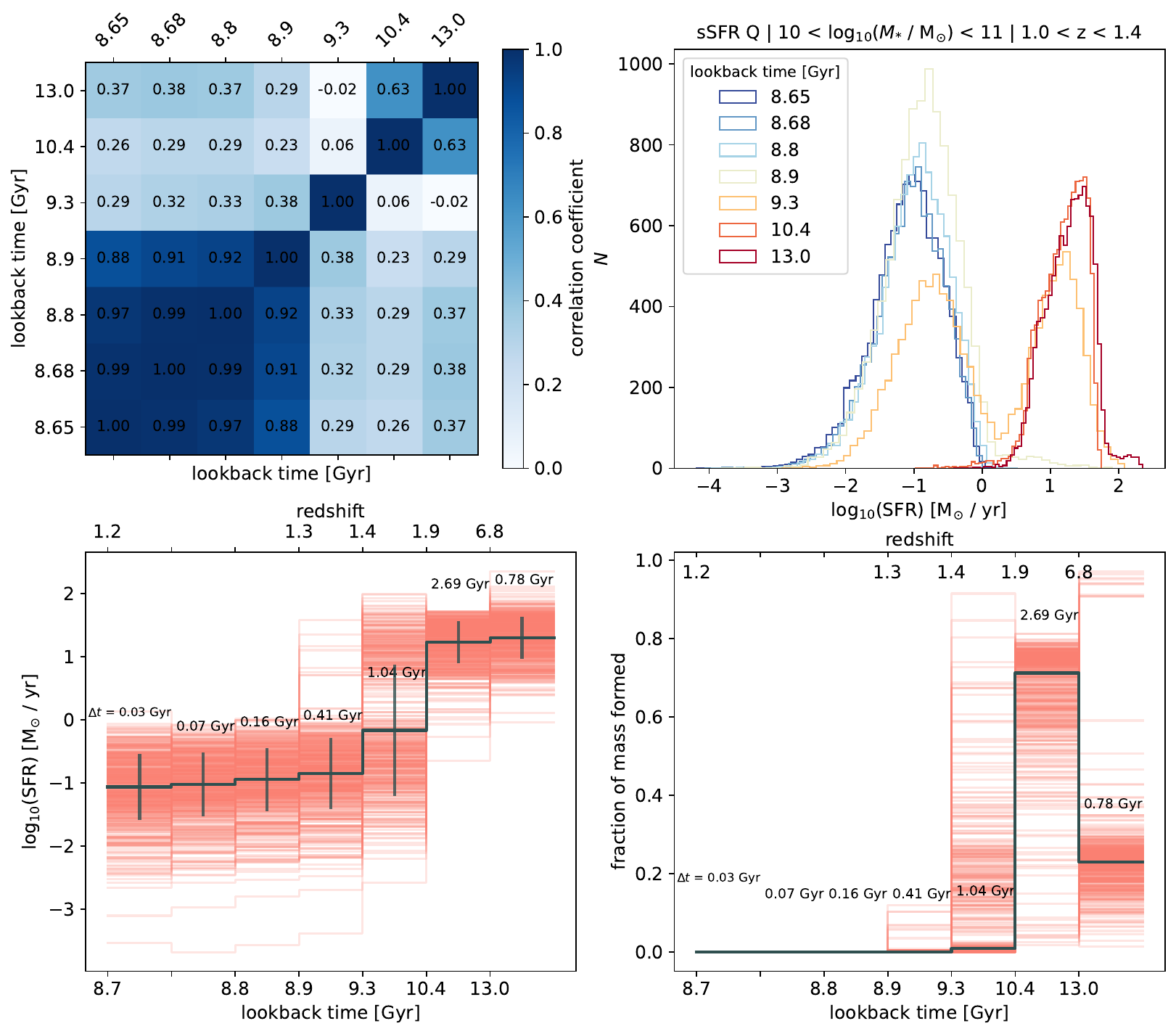}
    \caption{Same as Figure~\ref{fig:sf_massive}, but for the Q sample at $1.0<z<1.4$, with $10<\log_{10}(M_*/\mathrm{M}_\odot)<11$ and $\log_{10}(\text{sSFR}/\text{yr}^{-1})<-11$.}
    \label{fig:q_massive}
\end{figure*}

We present a detailed view of the star formation and stellar mass assembly of a galaxy population at $z\simeq1.2$ in Figures~\ref{fig:sf_massive} and \ref{fig:q_massive}. This redshift corresponds to the peak of the \pc\ redshift distribution \citep[see][]{thorp25b}, and it is also where the \pc\ SFRD peak occurs as we presented in Figure \ref{fig:sfrd}. We investigate a high ($10^{10}$--$10^{11}\,\mathrm{M}_\odot$) stellar mass range for the SF and Q subpopulations in the main body of this section, with a lower ($10^{9}$--$10^{10}\,\mathrm{M}_\odot$) stellar mass population discussed in Appendix \ref{app:sfh_lowmass}. Figures~\ref{fig:sf_massive} and \ref{fig:q_massive}, and the figures in Appendix \ref{app:sfh_lowmass}, all show the same four panels, with each figure showing the results for one of the subpopulations, as denoted above the top right panel. 

The top left panel is a $7\times7$ correlation matrix, corresponding to the seven-bin SFH. The rows and columns show the appropriate lookback time at $z \simeq 1.2$ for each SFH bin\footnote{The quoted lookback times in the figures are measured from $z=0$. In our assumed cosmology \citep{planck18}, redshift $z=1.2$ corresponds to a lookback time of $\sim8670$~Myr from $z=0$. It is also useful to consider the SFH in terms of lookback time from $z=1.2$. Expressed in this way, the eight bin edges defining the seven-bin SFH of a $z=1.2$ galaxy will be at $[0, 30, 100, 260, 660, 1690, 4350, 5120]$~Myr into its history.}. The matrix cells each contain the linear correlation coefficient between the SFR in the corresponding SFH bins, going back to the age of the Universe (i.e.\ spanning the lifetime of the subpopulation observed at $z \simeq 1.2$). The top right panel shows the (unnormalized) histogram of SFR in each of the $7$ SFH bins for the subpopulation. In the bottom left panel, we show the median SFR within the SFH bins of all galaxies in the subpopulation, alongside the individual SFRs of randomly selected draws from the relevant subpopulation. Recalling that the SFH bins are of different duration, we annotate the time interval represented by each bin, appropriately computed for $z\simeq 1.2$ where relevant. Finally, the bottom right panel illustrates the fraction of total stellar mass formed per SFH bin. This fraction is computed for the total stellar mass the galaxy has formed during its history, and not for the stellar mass remaining at its observed redshift.

For massive galaxies (Figures \ref{fig:sf_massive} and \ref{fig:q_massive}), the top left panels show that star formation 0.4~Gyr ago correlates with that at 1 Gyr ago at $r = 0.74$ for SF galaxies but only $r = 0.38$ for Q galaxies. The transition from active star formation to quiescence occurs within $\sim 1$ Gyr. Comparing the correlation matrices of SF and Q galaxies between Figures \ref{fig:sf_massive} and \ref{fig:q_massive}, we see that SF galaxies show a gradual decorrelation of SFR for more widely spaced bins. For the Q galaxies, we see a strikingly different behaviour with the most recent $\sim700$~Myr showing extremely strong correlation, indicating that once they have settled to a quiescent state, these galaxies hold steady at this reduced activity level for the remainder of their histories. 

The $\sim 1$ Gyr quenching time-scale is consistent with theoretical predictions for AGN feedback. The correlation structure rules out both instantaneous quenching (which would show sharp and complete decorrelation; e.g.\ \citealp{faisst17, zheng22, boselli22, park23_rapid, park24_rapid}) and slow strangulation alone (which would produce gradual decorrelation over $>2$~Gyr; e.g.\ \citealp{larson80, balogh00, bekki02, peng15, trussler20}). Our results on the correlation structure in the SFH of Q galaxies qualitatively agrees with the study of SFH within cosmological simulations by \citet{iyer20}, using the complementary power spectral density (PSD) approach. They find that in all simulations studied, Q galaxies show long time-scale correlation evidenced by elevated power at $>1$~Gyr time-scales, relative to SF galaxies.

The SFR distributions (top-right panels) reveal that SF galaxies maintain  $\gtrsim0.5$~dex scatter at all lookback times, indicating universal stochasticity independent of cosmic epoch. The distributions at earlier times appear very strongly constrained to $\text{SFR}>1\,\mathrm{M}_\odot\,\text{yr}^{-1}$.  A longer tail down to $\text{SFR}\simeq0.1\,\mathrm{M}\,\text{yr}^{-1}$ emerges at later times in the SFHs of these galaxies. The distributions in all lookback time bins have their mode at $\text{SFR}\simeq10\,\mathrm{M}_\odot\,\text{yr}^{-1}$. 

The median SFHs (bottom-left panels) show that SF galaxies in Figure \ref{fig:sf_massive} had sustained SFR $\sim 3$--30\,$\mathrm{M}_\odot\,\text{yr}^{-1}$ over their full lifetimes. While the median SFH of this subpopulation is fairly flat, for any given galaxy a highly variable SFH is possible, as evidenced by the bin-to-bin correlation plot in the top left panel of Figure \ref{fig:sf_massive}. Conversely, the Q galaxies in Figure~\ref{fig:q_massive} show a complete cessation ($\text{SFR}\lesssim 0.1\,\mathrm{M}_\odot\,\text{yr}^{-1}$) for the past $\sim700$~Myr, preceded by a phase of vigorous star formation ($\gtrsim10\,\mathrm{M}_\odot\,\text{yr}^{-1}$) during the first $\sim3$~Gyr of their existence. Interestingly, there appears in both the top right and bottom left panels of Figure \ref{fig:q_massive} a transitional epoch lasting $\sim1$~Gyr between 700 and 1700~Myr into their past where $\sim50$ percent of the Q population has ceased rapid star formation ($\text{SFR}\simeq0.1\,\mathrm{M}_\odot\,\text{yr}^{-1}$), whilst the remainder continue forming stars at close to their peak rates ($\sim10\,\mathrm{M}_\odot\,\text{yr}^{-1}$). Individual galaxy tracks reveal that a fraction of nominally quiescent galaxies retain residual star formation as high as $\sim0.5\,\mathrm{M}_\odot\,\text{yr}^{-1}$ in recent epochs, suggesting incomplete quenching or minor rejuvenation events.

The bottom right panels of Figures \ref{fig:sf_massive} and \ref{fig:q_massive} show the fraction of the stellar mass formed in each of the SFH bins. While it is not surprising that the distribution peaks in the penultimate bin corresponding to the longest duration of $2.69$~Gyr, the SF/Q classification is based on sSFR in the two most recent SFH bins, so it is informative to compare and contrast these distributions for the two subpopulations. Looking at the bottom right panel of Figure \ref{fig:sf_massive}, we see that there are some SF galaxies that formed up to $\sim10$ percent of their stellar mass in very short $\sim30$--70~Myr time-scales ($\lesssim1$ percent of their lifetimes) recently in their history, indicative of the considerable ongoing activity at $z\simeq1.2$. We see that massive Q galaxies typically formed $\sim 95$ percent of their stellar mass more than 2 Gyr before they are observed, whereas massive SF galaxies only formed $\sim 50$--$60$ percent of their mass over the same period. 

The behaviour of the lower mass galaxies discussed in Appendix \ref{app:sfh_lowmass} remains qualitatively the same as for the more massive subpopulation discussed above, albeit with some quantitative differences indicating a slower and more complex quenching transition. 

\subsection{The \pc\ model as a prior on SFH}
\label{subsec:sfh_discussion}

\begin{figure}
    \centering
    \includegraphics[width=\linewidth]{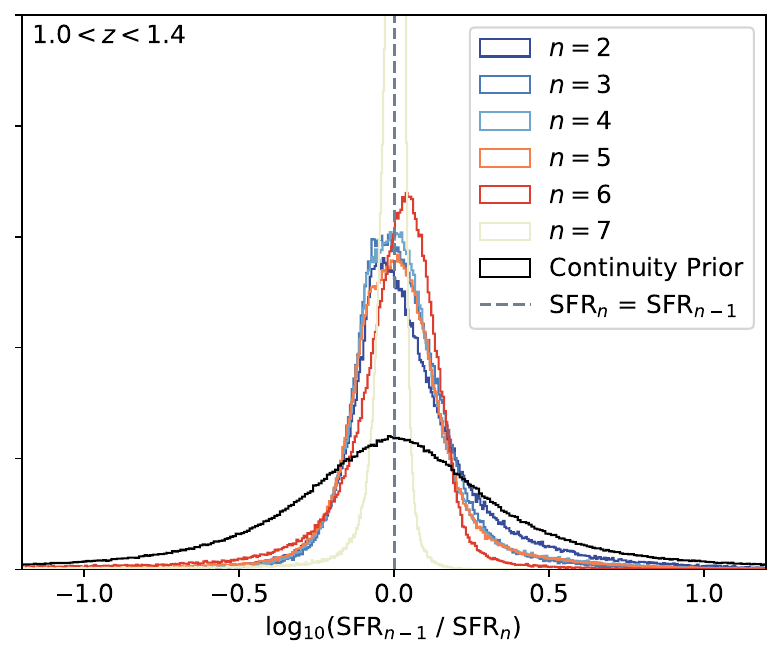}
    \caption{Distributions of SFR ratios between the SFH bins for all the galaxies in $1.0 < z < 1.4$. SFH bins~$n=1$ and $n=7$, respectively correspond to the most recent and earliest epochs of the SFH. The SFR ratio plotted is therefore the ratio of a more recent bin to the immediately preceding bin. The dashed vertical line shows an SFR ratio of one. The black histogram shows the Student's $t$ continuity prior from \citet{leja19_sfh}. All histograms are normalized to integrate to one.}
    \label{fig:sfr_ratios}
\end{figure}

We now turn to the implications of the population-level SFR distributions seen in the seven-binned SFH parametrization. As with the \texttt{Prospector}-$\alpha$ prescription \citep[see][]{leja17, leja19_sfh, leja19}, our SPS model has six free parameters defining the ratios of the SFR in adjacent bins of the seven-bin SFH. We define the most recent bin in the SFH (corresponding to the most recent 30~Myr of a galaxy's life) as bin $n=1$, and the bin with the furthest lookback time (corresponding to the first 15 percent of the Universe's life, viewed from a galaxy's redshift) as bin $n=7$. The SFR ratios are denoted as $\Delta\log_{10}(\text{SFR})_n=\log_{10}(\text{SFR}_{n-1}/\text{SFR}_{n})$ where $n=2,3,\dots,7$. 

\cite{leja19_sfh} introduce a `continuity prior' as a plausible distribution for these ratios. The continuity prior favours smoother transitions between adjacent bins by sampling these ratios from a Student's $t$-distribution. The heavy-tailed distribution is sufficiently flexible to allow for diverse SFHs with bursts or fast quenching, and is neutral with regard to the direction of change of the SFR between bins -- i.e.\ it is symmetric with median $\Delta\log_{10}(\text{SFR})=0$. Variations on this form to allow for more dramatic starbursts \citep[e.g.][]{tacchella22a}, bins that are not fixed in time \citep[e.g.][]{leja19_sfh, suess22}, and finer time binning \citep[e.g.][]{wang25} have all been explored in the literature.

While using the same non-parametric binning framework to parametrize our SFH, we do not impose a specific population-level prior such as the continuity prior on the \pc\ generative model. During training, the model is free to calibrate the SFR ratios between adjacent bins to achieve an optimal fit to the COSMOS2020 training data. Therefore, the distribution of SFR ratios in the trained model reveal insights about the SFH distributions of the galaxy population for which the COSMOS2020 training data is representative. Moreover, the learned joint distribution over the SFR ratios can be used as a data-driven prior in an SED fitting context, as shown by \citet{thorp24}.

Figure \ref{fig:sfr_ratios} shows the learned distribution of the SFR ratios, plotted for all mock galaxies in the redshift range $1.0<z<1.4$ (i.e. at the peak of our SFRD), compared with the continuity prior from \citet{leja19_sfh}. It should be borne in mind that this plot is produced by marginalizing over all the other parameters in the population model, including stellar mass. We see that the \pc\ model's learned SFR ratio distributions differ systematically from the continuity prior. The model learns substantially narrower distributions for all six SFR ratios, indicating that galaxy SFHs follow more constrained evolutionary pathways than the continuity prior assumes. 

Strikingly, the ratio between the penultimate and final bins peaks sharply at unity, reflecting the fundamental limitation that the oldest stellar populations contribute minimally to integrated galaxy SEDs. The impact of this stellar population is even more difficult to capture with photometry alone. The model appropriately infers that these poorly-constrained SFRs deep in the history of the galaxy should closely track the subsequent epoch's SFR (i.e.\ the model extrapolates SFH conservatively when there is little information in the data). It is worth emphasizing that this result is a statement about the lack of information in the COSMOS2020 data about SFR of the oldest stellar populations at these early times; it does not imply that the actual SFH of galaxies are not changing during this epoch. Moving to later cosmic times, the distributions progressively broaden, though never approaching the width of the continuity prior. 

Considering the $n = 6\to5$ ratio in this redshift range, the distribution of $\Delta\log_{10}(\text{SFR})_6$ is peaked at a positive value of $\sim0.1$, indicating rising SFH towards the peak of the SFRD. The SFR ratio distributions of the most recent bins ($n = 3\to2$ and $n = 2\to1$) peak at lower values of $\sim{-0.1}$; examination of the SF/Q contributions in these bins show that this decline in SFH is driven by galaxies that were already quenched at these epochs, as seen in Figure \ref{fig:q_massive} (and more strongly for the lower-mass Q galaxies in Figure \ref{fig:q_lowmass}). Further, these distributions exhibit positive skewness -- their tails extend preferentially toward positive values, corresponding to rising SFH trajectories where earlier epochs have lower SFRs than later ones. The strength of this skewness grows as galaxies approach the most recent bins in their SFH. 

As noted above, these results are conditioned on the time binning approach of \citet{leja19_sfh, leja19}. Any binning approach is necessarily a lossy representation of the SFH of the galaxy population, not only because of averaging effects but because any specific dataset will exhibit information loss on the underlying SFHs of galaxies in the sample due to effects such as outshining \citep{wang25}. It is necessary to balance the complexity of the non-parametric SFH representation with the information content in the data. In future calibrations of \pc\ we will explore the potential for optimal design of the number and positioning of bins in non-parametric SFH modelling using these data-driven approaches.

\section{Discussion}
\label{sec:discussion}

We have presented a multi-faceted picture of stellar mass assembly and star formation at $z < 3.5$ as learned by the \pc\ generative model. Here we discuss some implications of our results for the standard galaxy evolution paradigm, also connecting to the role of AGN.

\subsection{Cosmic SFRD in context}

The primary difference between our results and the standard `cosmic noon' picture of star formation in the Universe is the shift in the peak of the SFRD to approximately 2 Gyr later cosmic times (Figure \ref{fig:sfrd}). With comparable peak amplitudes to \cite{madau14}, this is not a change in overall star formation, but rather a temporal redistribution of when that star formation occurred. One interpretation of the result is that the deep IR selection on which \pc\ is calibrated has captured dust-obscured star formation that was under-represented in earlier compilations. Many studies contributing to the \cite{madau14} synthesis relied on UV LFs with dust corrections that may have been incomplete, particularly at $z\simeq 1.0$--$1.5$ where dust obscuration is significant \citep[e.g.][]{zavala21}. The agreement between \pc\ and the CIB measurements from \cite{chiang25} supports this interpretation -- both methods that better account for dust-reprocessed light find the peak at lower redshift than UV-dominated surveys. 

\begin{figure}
    \centering
    \includegraphics[width=\linewidth]{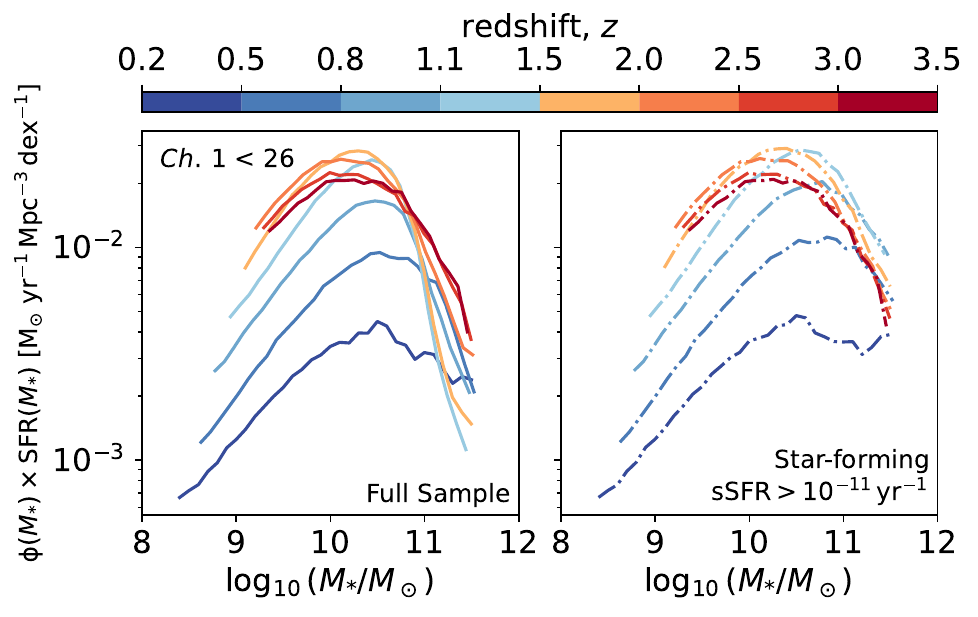}
    \caption{Mass dependent contribution to the SFRD, with the same redshift binning as Figure \ref{fig:smf_separate}. Estimated by multiplying the median star-forming sequence from \citet{thorp25b} with the SMF from this work. \textbf{Left:} Full sample of mock galaxies. \textbf{Right:} SF sample based on sSFR selection. Plotted curves are truncated at the completeness limit described in Section \ref{subsec:data}.}
    \label{fig:SFSxSMF}
\end{figure}

The shift to a later peak while maintaining a similar SFRD amplitude implies that the cosmic star formation history may be more extended than the standard picture -- rather than peaking sharply at $z\simeq2$, there is a broader plateau of high star formation extending from $z\simeq2$ down to $z\simeq1$. To investigate the mass scales driving the evolution of the SFRD, we show in Figure \ref{fig:SFSxSMF} the SFRD contribution of galaxies as a function of stellar mass and redshift. This is obtained (as described in e.g.\ \citealp{leja15}) by weighting the \pc\ star-forming sequence from \citet{thorp25b} by the SMF presented in Section \ref{sec:smf} of this work. The left-hand panel shows all galaxies, while the right-hand panel selects out the star-forming sample in order to clarify the trends within that subpopulation. Looking at the right-hand panel of Figure \ref{fig:SFSxSMF}, we see that the mass scale with peak contribution evolves with redshift: at $z\simeq3.5$ the SFRD contribution peaks at $\sim10^{10}\,\mathrm{M}_\odot$, with the peak mass scale evolving very gradually upwards to $\sim10^{10.5}\,\mathrm{M}_\odot$ by the peak of the SFRD at $z\simeq1.3$, and further to almost $\sim10^{11}\,\mathrm{M}_\odot$ by $z\simeq0.5$ (although the peak mass scale at $0.2<z<0.5$ is less well-defined). We find that a broad range of masses contributes to the SFRD at $z\gtrsim1.5$, including many of moderate stellar masses $\sim10^{10}\,\mathrm{M}_\odot$, with the contribution of higher-mass galaxies growing as the peak of the SFRD is passed at $z\simeq1.3$. 

These results align with a scenario where the quenching efficiency is lower than the standard picture, driving the extended peak of the SFRD seen in Figure \ref{fig:sfrd}. The high estimated contribution of $\sim10^{10}\,\mathrm{M}_\odot$ galaxies at $z\gtrsim1.5$ may be driven by dusty star-forming galaxies, expected to be well covered by the deep IR selection used in the calibration of \pc\, and enabled by COSMOS2020. As previously argued, such galaxies, likely to have been missed by shallower or UV-selected surveys, could be responsible for the elevated SFRD at $z\gtrsim1.5$ that we see in Figure \ref{fig:sfrd}. 

These estimates also bear on the question of how effective integration limits differ between our approach and traditional LF-based SFRD measurements. Analyses such as \cite{madau14} integrate a fitted luminosity function down to a minimum luminosity defined relative to a characteristic value $L_*$ ($0.03\times L_*$ for \citealp{madau14}), while our approach sums SFRs for all galaxies above our redshift-dependent mass completeness limit. While this does mean that the effective SFR range is broader at lower $z$ (since the flux limit corresponds to progressively fainter rest frame luminosities at lower $z$), the differential contribution from galaxies near or below the completeness limit is bounded at a few per cent. This is insufficient to account for the $\Delta z\simeq 0.6$ shift in the SFRD peak.

Whilst the depth of COSMOS2020 allows for a high degree of mass-completeness down to $\sim10^{9}\,\mathrm{M}_\odot$, we can estimate from Figure \ref{fig:SFSxSMF} the potential SFRD contribution that is missed in our analysis due to SF galaxies below our mass completeness limit. To do this, we fit a double power law to each of the curves in the right hand panel of Figure \ref{fig:SFSxSMF}, and use this to extrapolate below our completeness limit to estimate the contribution in this low-mass tail. At $z\lesssim1.5$, we estimate around 3--5 percent of the total SFRD to be below our completeness limit. At $z\simeq3$ we conservatively estimate that up to $\sim 10$ percent of the total SFRD is lost. We will be able to validate these estimates in future calibrations of \pc\ using COSMOS2025 \citep{shuntov25} which includes MIR \textit{JWST} data.

While noting that beyond $z\simeq 3.5$, the \pc\ results are extrapolating somewhat beyond the regime well-constrained by COSMOS2020 number counts, we find that the \pc\ SFRD flattens out relative to the standard \citet{madau14} expectations or a constant star formation efficiency model \citep[e.g.][]{mason15, tacchella18, bouwens21, harikane22} at higher redshifts. It is noteworthy that recent high-redshift ($z\gtrsim8$) measurements of the SFRD are more consistent with a flatter evolution in the number densities \citep{harikane23, donnan23, bouwens23, finkelstein24}. In the future we will attempt to constrain this regime better with deeper calibrations of \pc\ to make closer contact with the high-redshift regime probed by these studies.

\subsection{{Dust-obscured star formation}}

\begin{figure}
    \centering
    \includegraphics[width=\linewidth]{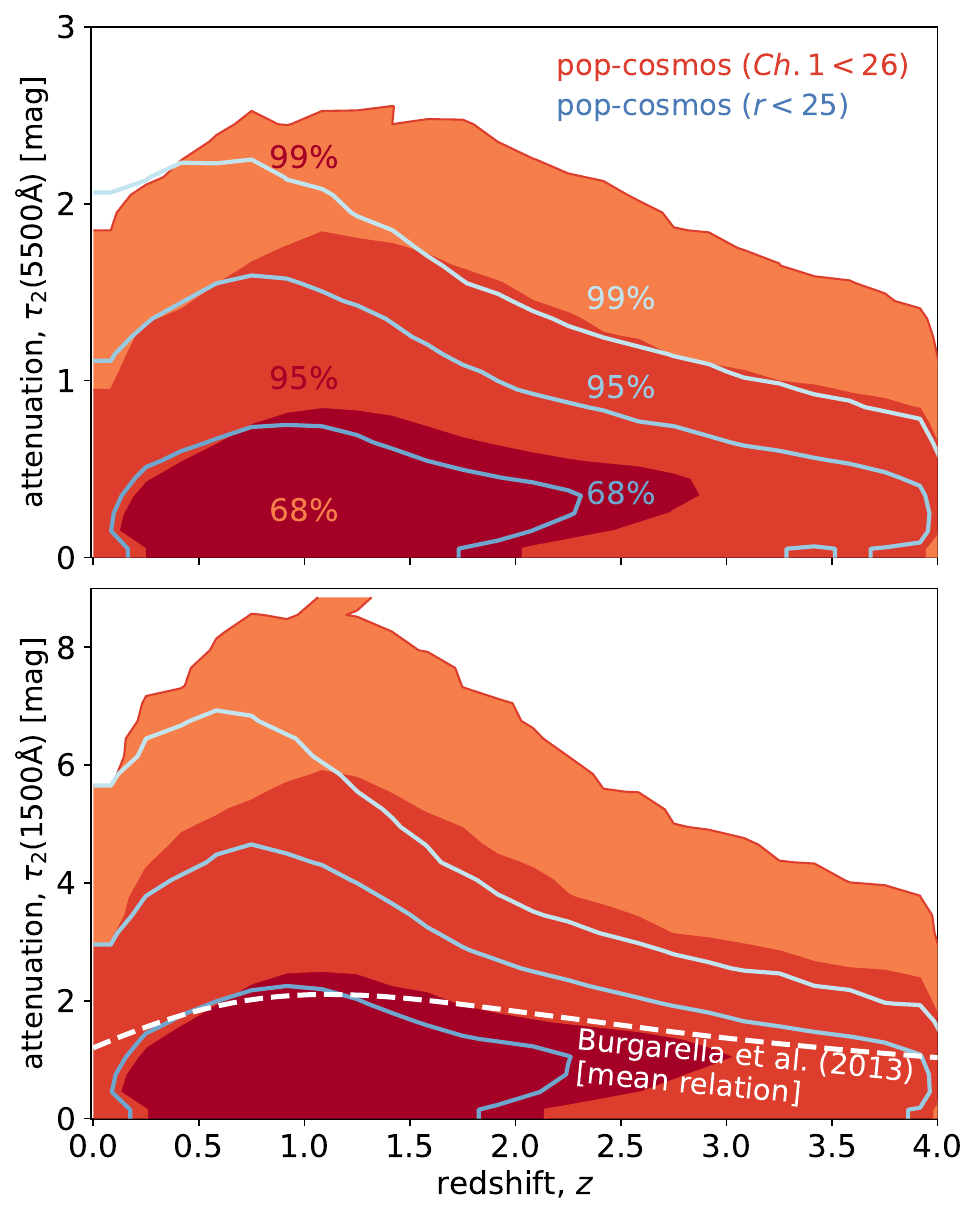}
    \caption{{Joint distribution of diffuse dust attenuation and redshift for \pc. \textbf{Top:} Optical attenuation (at $5500$~\AA) in magnitudes. \textbf{Bottom:} Far UV attenuation (at $1500$~\AA) in magnitudes. Red filled contours show the $\textit{Ch.\,1}<26$ mock galaxy catalogue upon which our main analysis is based. Light blue open contours show the companion $r<25$ mock catalogue from \citet{thorp25b}. Contours are drawn to contain 68, 95, and 99 percent of the population. The white dashed line shows the mean UV attenuation vs.\ redshift relation from \citet{burgarella13}.}}
    \label{fig:dust_redshift}
\end{figure}

{To further investigate the role of dust in our SFRD inference, we explore here the correlation between dust attenuation and redshift that has been learned by the \pc\ model. We refer the reader to \citet{alsing24}, \citet{thorp25b}, and \citet{petri25} for further investigations on the correlations between attenuation, SFR, and stellar mass. In Figure \ref{fig:dust_redshift}, we show the joint distribution of diffuse dust attenuation, $\tau_2$, and redshift, $z$, for the mass-complete $z<4$ \pc\ mock catalogue. We show the attenuation at both 5500~\AA\ ($\sim V$-band, the fiducial wavelength for the $\tau_2$ parameter in \texttt{FSPS}), and 1500~\AA\ (the far-UV, as assumed in e.g.\ \citealp{madau14}). We compute the far-UV attenuation using the \citet{noll09} modification of the \citet{calzetti00} attenuation curve, with UV bump strength correlated with slope following \citet{kriek13}.}

{From Figure \ref{fig:dust_redshift}, we can see that the prevalence of dusty galaxies is highest around the peak of our SFRD at $z\simeq1.3$. This being the peak redshift for dust obscuration aligns well with the far-IR analysis by \citet{burgarella13} using \textit{Herschel} data, and the SED fitting analysis by \citet{cucciati12}, which together form a large part of the $z\lesssim4$ compilation of attenuation estimates in \citet{madau14}. We overplot on Figure \ref{fig:dust_redshift} the fitting function for the mean far-UV attenuation as a function of $z$ provided by \citet{burgarella13}, showing its correspondence with the \pc\ result. One thing that is apparent from these results is that the scatter in attenuation values at any given redshift is substantial, likely far larger than can be well represented or corrected for with a single mean relation.} Moreover, because the most heavily star-forming galaxies are also typically the most heavily attenuated, a luminosity-weighted effective UV attenuation for the full population would be expected to peak more sharply around $z\simeq1$--1.5 than the mean relation of \citet{burgarella13}. This would amplify the difference between our directly-computed SFRD and UV-based estimates that apply a single mean attenuation correction to the luminosity function.

{As a further test, we also investigate the dust--redshift correlation in the optically selected ($r<25$) mock catalogue from \citet{thorp25b}. The distribution of these galaxies in $\tau_2$ vs.\ $z$ space is indicated on Figure \ref{fig:dust_redshift} with open blue contours. As expected, we see that the population probed by the $\textit{Ch.\,1}<26$ selection is substantially dustier (see also the results in \citealp{petri25}), especially at $z\gtrsim1$, with the most heavily attenuated galaxies in the $r<25$ population being found at $z\lesssim1$. Whilst these tests cannot show definitively that a $\textit{Ch.\,1}<26$ selection will identify \emph{all} dusty star-forming galaxies, it does indicate that such a selection will be substantially more complete than a deep optical sample, and that highly-reddened galaxies are well represented over the redshift range we have considered in this work.}

\subsection{AGN activity and the transition to quiescence}

\begin{figure}
    \centering
    \includegraphics[width=0.9\linewidth]{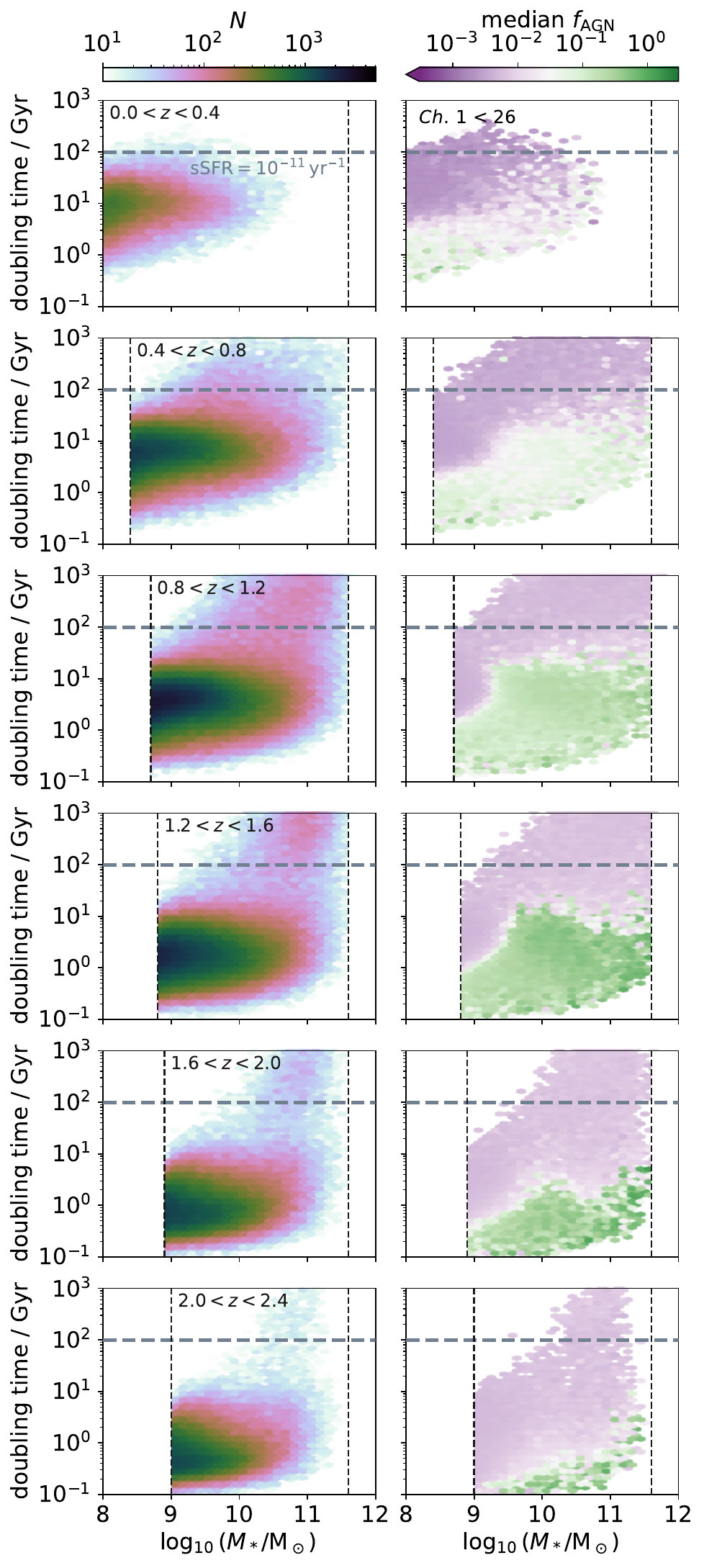}
    \caption{Stellar mass doubling time-scale (inverse sSFR; see, e.g.\ \citealp{bower17}) vs.\ stellar mass in bins of redshift. \textbf{Left:} Cells shaded based on galaxy count. \textbf{Right:} Cells shaded based on median AGN bolometric luminosity fraction, $f_\text{AGN}$. The dashed gray line indicates a doubling time of $100$~Gyr, equivalent to our SF/Q boundary of $\text{sSFR}=10^{-11}\,\text{yr}^{-1}$. The vertical dashed lines show the mass completeness limit and upper mass threshold described in Section \ref{subsec:data}.}
    \label{fig:inverse_ssfr}
\end{figure}

We now leverage the interpretability built into the \pc\ generative model to explore the physical quenching mechanisms at play. The model has learned non-linear relationships between star formation, mass assembly, and signatures of AGN activity during its training process, within its SPS parametrization. Figure \ref{fig:inverse_ssfr} maps the \pc\ mock catalogue in stellar mass growth rate (inverse sSFR) vs.\ stellar mass space across cosmic time \citep[inspired by][]{bower17}. The left panels reveal a bimodal distribution: a star-forming peaking at sequence at $1/\text{sSFR}\simeq1$--$10$ Gyr (doubling times of $1$--$10$ Gyr) and a quiescent cloud at $1/\text{sSFR}> 100$ Gyr (essentially no ongoing growth) that emerges from $z\lesssim2$. Most distinctly at $0.8\lesssim z \lesssim2.0$, we see a clear transitional region connecting the SF and Q clouds, with doubling times $10$--100~Gyr, and a characteristic stellar mass of $\sim10^{10.5}\,\mathrm{M}_\odot$, which appears to mark a threshold for the quenching transition \citep[see e.g.][]{bower17}. This stellar mass scale also coincides closely with the point where the slope of the SFR vs.\ stellar mass relation in the star-forming sequence flattens \citep[e.g.][]{whitaker12, whitaker14, lee15, schreiber15, tasca15, leja22, popesso23, thorp25b}, marking a characteristic mass scale where a significant quiescent population emerges.

The relationships learned by \pc\ show that AGN activity traces the edge of this quenching transition. The right panels of Figure \ref{fig:inverse_ssfr} reveal a striking correlation; the AGN bolometric luminosity fraction ($f_\mathrm{AGN}$) at $0.8\lesssim z \lesssim 1.6$ peaks as galaxies approach the lower edge of the transitional region, reaching median values of $f_\mathrm{AGN} \simeq 0.1$ to $0.5$ ($10$--$50$ percent of bolometric luminosity) at doubling times of $\sim3$--10~Gyr, and stellar masses $\gtrsim10^{10}\,\mathrm{M}_\odot$. This `AGN ridge' at $1/\text{sSFR}\simeq 3$--$10$ Gyr marks a critical point -- galaxies nearing this ridge are still forming stars but at suppressed rates relative to the main sequence. Once galaxies cross the ridgeline at growth scales of $\sim10$~Gyr, and begin transitioning towards the fully quiescent region ($1/\text{sSFR}> 100$ Gyr), the median $f_\text{AGN}$ drops dramatically, and for quiescent galaxies is typically $< 10^{-2.5}$ (i.e.\ less than 0.3 percent of bolometric luminosity). This is one or two orders of magnitude lower than in the objects at the edge of the ridge. 

The AGN signature we model in \pc\ with the $f_\text{AGN}$ parameter is an IR-bright dust torus \citep{nenkova08i, nenkova08ii}, corresponding to an efficiently accreting black hole. It is interesting to consider the degree of overlap between this signature and other observables of this rapid accretion phase. There is a degree of overlap between IR-bright and X-ray bright AGN \citep[e.g.][]{juneau13, lamassa19, carroll21}, with these two signatures both tracing high accretion-rate `quasar-mode' systems. Optical emission-line diagnostics can also trace this phase \citep{juneau13}, and detailed multi-wavelength SED fits have been used to separate different modes of AGN activity \citep[e.g.][]{marshall22, thorne22a, zou22, martinez24}. In our previous work \citep{thorp25b}, we found that X-ray-detected galaxies in COSMOS2020 \citep{civano16} had preferentially higher $f_\text{AGN}$ estimates in the IR, corroborating this picture.

The AGN ridge identifies galaxies caught in the act of quenching. The baryonic redistribution from AGN feedback is expected to postdate the rapidly-accreting phase discussed above, with the feedback phase corresponding to lower gas accretion rates and eventually quenching star formation \citep{bower17}. The transitional objects with elevated $f_\mathrm{AGN}$ and intermediate sSFR represent the $\sim 1$~Gyr phase identified in Section \ref{sec:sfh} where galaxies start migrating from the star-forming main sequence to quiescence. The sequence appears to proceed as follows: as galaxies approach a stellar mass scale of $\sim10^{10.5}$ M$_\odot$, $f_\mathrm{AGN}$ rises, suggestive of rising AGN activity. It is notable that in still star-forming galaxies, this AGN signature is strongly correlated with a high SFR, indicating that both are driven by the same underlying mechanism (presumably the infall of gas). AGN feedback reduces sSFR from $\sim1\,\text{Gyr}^{-1}$ to $\sim0.1\,\text{Gyr}^{-1}$, finally driving galaxies over the ridge and towards full quiescence. Once galaxies move into the transitional region with $\text{sSFR}<0.1\,\text{Gyr}^{-1}$, reduced gas supply causes $f_\mathrm{AGN}$ to decline as accretion rate slows and the AGN's IR bright phase ends \citep[see e.g.][]{padovani17}. Our results are consistent with the  merger-driven picture of AGN quenching  investigated by \cite{davies22}. By doing controlled experiments using `genetically modified' simulations, they find that the onset of a merger in a galaxy with stellar mass $10^{10.5}\,\mathrm{M}_\odot$ drives an initial rise in the SFR and accretion rate, followed by quenching over a $\sim1$~Gyr time-scale due to AGN activity, leaving the galaxy gas-poor.

The width of the transition from star-forming to quiescent in $1/\text{sSFR}$ space (spanning $\sim 1$ dex from $10$ to $100$ Gyr), combined with the width of the quenching transition ($\sim1$~Gyr) identified in Section \ref{sec:sfh}, constrains the quenching time-scale. If galaxies traverse this region in $\sim 1$ Gyr, the implied migration rate is $\mathrm{d}[\log_{10}(1/\text{sSFR})]/\mathrm{d}t \simeq 1~\text{dex}\;\text{Gyr}^{-1}$. The concentration of high $f_\mathrm{AGN}$ values in the approach to this region, and tight coupling between transitioning AGN and star formation activity, suggests that AGN feedback operates in a critical regime where star formation-driven outflows balance gas accretion by the AGN \citep{bower17}.

\section{Conclusions}
\label{sec:conclusions}

We have leveraged the \pc\ generative model \citep{alsing24, thorp25b} to investigate stellar mass assembly and star formation history across cosmic time, providing new constraints on the cosmic SFRD and the star formation histories of galaxy populations. The approach we have taken offers a new window into these properties, being based on the distributions of SPS parameters learned from a population-level analysis of a deep photometric survey. This avoids the challenges inherent in inferring global quantities from SED fits of individual galaxies, especially from heterogeneous surveys, and provides a more complete view of galaxy evolution through directly learning a joint distribution over a sophisticated SPS parametrization at the population level.

Our analysis reveals that the cosmic SFRD peaks at $z \simeq 1.3$, approximately $\Delta z \simeq 0.6$ later than the canonical \cite{madau14} estimate, while having a similar normalization. We attribute this shift to the greater sensitivity of our deep IR-selected COSMOS2020 calibration to dust-obscured star formation in intermediate-mass galaxies at $z \simeq 1$--$1.5$, which were under-represented in earlier UV-dominated compilations. The agreement with recent CIB measurements \citep{chiang25} independently confirms that a more complete census of cosmic star formation extends to later times than previously recognized.

Our systematic comparison of sSFR- and colour-based galaxy classification into star-forming and quiescent populations reveals that $NUVrJ$ selection suffers from up to 20 percent contamination by dusty star-forming galaxies with sSFR $>10^{-11}$ yr$^{-1}$, altering the inferred low-mass slope of the quiescent SMF. In particular, we find a negligible density of low-mass ($\lesssim10^{9.5}\,\mathrm{M}_\odot$) quiescent galaxies at $z\simeq1$. Analysis of the full seven-bin non-parametric SFHs demonstrates that massive galaxies ($10^{10}$--$10^{11}$ M$_\odot$) undergo quenching on $\sim 1$ Gyr time-scales, with AGN activity peaking (median $f_\text{AGN}\simeq10$--50~percent) as galaxies start to enter the transition between star-forming and quiescent states. The correlation structure of the SFHs -- gradual decorrelation for star-forming galaxies versus sharp transitions for quiescent galaxies -- provides direct evidence for distinct quenching mechanisms operating above the critical mass scale $\sim 10^{10.5}$ M$_\odot$. These results also align with the recent finding by \cite{lucie25} using the FLAMINGO simulations \citep{schaye23} that, regardless of redshift, AGN feedback  
most efficiently redistributes baryons when haloes reach a critical mass scale of $M_\mathrm{200m} \simeq 10^{12.8}$ M$_\odot$, which corresponds to a stellar mass scale of $\sim 10^{11}$ M$_\odot$ using the FLAMINGO stellar-to-halo mass relation. 

These population-level constraints from \pc\ offer a new window into galaxy evolution, demonstrating the power of generative models trained on deep photometric surveys to reveal the diversity and complexity of stellar mass assembly across cosmic time. In future work we will investigate the robustness of our conclusions to dust emission \citep[see e.g.][]{draine07, jones17_themis} and AGN modelling \citep[see e.g.][]{temple21, wang24_agn} assumptions, and leverage data from further in the IR (e.g.\ the \textit{JWST F770W} photometry in COSMOS2025{, which will be sensitive to AGN-heated dust at higher-$z$ than \textit{Spitzer} IRAC \textit{Ch.\,1} and \textit{2}}; \citealp{shuntov25}) in an even deeper calibration of \pc\ to trace the relationship between star formation and AGN activity at higher-$z$. Moreover, we plan to use state-of-the-art, flux-limited spectroscopic samples, such as DESI \citep{hahn23_bgs, adame24, abdul25} at low-$z$ and MOONRISE \citep{maiolino20} at high-$z$, within our calibration data to provide complementary information about galaxy evolution. Further, we will validate the self-consistency of the different galaxy evolution trends learned by \pc\ through a detailed comparison with the COLIBRE simulation suite \citep{schaye25, chaikin25}. 

\section*{Acknowledgements}
% https://credit.niso.org
We outline the different contributions below using keywords based on the Contribution Roles Taxonomy (CRediT; \citealp{brand15}).
\textbf{SD:} conceptualization; formal analysis; visualization; software; investigation; writing -- original draft, review \& editing.
\textbf{HVP:} conceptualization; methodology; visualization; investigation; validation; writing -- original draft, review \& editing; supervision; project administration; funding acquisition.
\textbf{ST:} data curation; methodology; software; validation; visualization; writing -- original draft, review \& editing.
\textbf{DJM:} methodology; validation; writing -- original draft, review \& editing.
\textbf{GJ:} software.
\textbf{JA:} methodology; writing -- review.
\textbf{BL:} methodology; writing -- review \& editing.
\textbf{JL:} validation; writing -- review \& editing.

We thank Anik Halder, Madalina Tudorache and Benedict Van den Bussche for useful discussions and collaboration. We thank John Weaver and Christian Kragh Jespersen for useful correspondence about COSMOS2020, and Ben Johnson for his SPS-related insights. We are grateful to Suchetha Cooray, Andrew Pontzen, Sandro Tacchella, and Risa Wechsler for helpful conversations. {We thank the referee for their helpful comments and detailed reading of the manuscript, which has strengthened the paper.}

SD, HVP, ST and JA have been supported by funding from the European Research Council (ERC) under the European Union's Horizon 2020 research and innovation programmes (grant agreement no.\ 101018897 CosmicExplorer). This work has been enabled by support from the research project grant `Understanding the Dynamic Universe' funded by the Knut and Alice Wallenberg Foundation under Dnr KAW 2018.0067. HVP was additionally supported by the G\"{o}ran Gustafsson Foundation for Research in Natural Sciences and Medicine. BL was supported by the Royal Society through a University Research Fellowship. HVP and DJM acknowledge the hospitality of the Aspen Center for Physics, which is supported by National Science Foundation grant PHY-1607611. This research was supported in part by Perimeter Institute for Theoretical Physics. Research at Perimeter Institute is supported by the Government of Canada through the Department of Innovation, Science and Economic Development and by the Province of Ontario through the Ministry of Research, Innovation and Science.

This research utilized the Sunrise HPC facility supported by the Technical Division at the Department of Physics, Stockholm University. This work was performed using resources provided by the Cambridge Service for Data Driven Discovery (CSD3) operated by the University of Cambridge Research Computing Service (\url{www.csd3.cam.ac.uk}), provided by Dell EMC and Intel using Tier-2 funding from the Engineering and Physical Sciences Research Council (capital grant EP/T022159/1), and DiRAC funding from the Science and Technology Facilities Council (\url{www.dirac.ac.uk}).

%%%%%%%%%%%%%%%%%%%%%%%%%%%%%%%%%%%%%%%%%%%%%%%%%%
\section*{Data Availability}

The mock galaxy catalogues based on the \citet{thorp25b} \pc\ model are publicly available on Zenodo \citep{thorp25_mock}. We have used the $\mlim<26$ mock catalogue (\verb!mock_catalog_Ch1_26.h5!) from an updated v1.1.0 of the Zenodo record that incorporates the rest-frame $NUVrJ$ photometry we introduced in this work. The upper limits on stellar mass and SFR are set in part based on the \pc\ posterior samples from SED fits to COSMOS2020 galaxies, published on Zenodo \citep{thorp25_mcmc}. We have used the posterior quantiles (\verb!mcmc_summaries.h5!) from v2.1.1 of the Zenodo record. The core \pc\ code \citep{alsing24, thorp24, thorp25b}, which can be used to generate mock catalogues, is available on GitHub: \url{https://github.com/Cosmo-Pop/pop-cosmos}. We have made use of the COSMOS2020 SMF function constraints associated with \citet{weaver23}, using the publicly available Zenodo record \citep{weaver23_zenodo}.



\bibliographystyle{mnras}
\bibliography{pop-cosmos} 

%%%%%%%%%%%%%%%%%%%%%%%%%%%%%%%%%%%%%%%%%%%%%%%%%%

%%%%%%%%%%%%%%%%% APPENDICES %%%%%%%%%%%%%%%%%%%%%

\appendix

\section{Star Formation Histories of Lower-Mass Galaxies}
\label{app:sfh_lowmass}

\begin{figure*}
    \centering
    \includegraphics[width=0.95\linewidth]{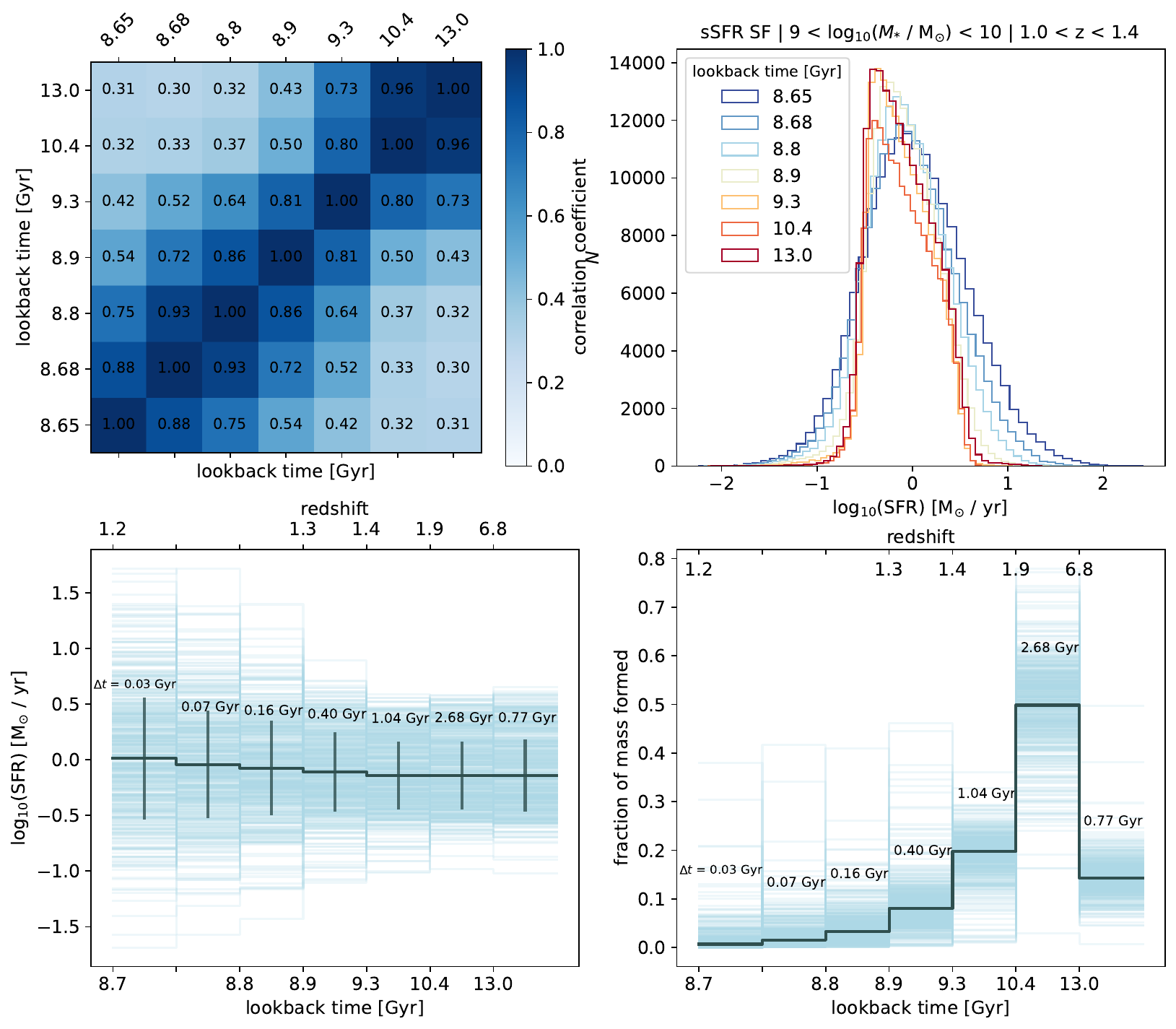}
    \caption{Same as Figure~\ref{fig:sf_massive}, but for a lower stellar mass range of $9<\log_{10}(M_*/\mathrm{M}_\odot)<10$.}
    \label{fig:sf_lowmass}
\end{figure*}

\begin{figure*}
    \centering
    \includegraphics[width=0.95\linewidth]{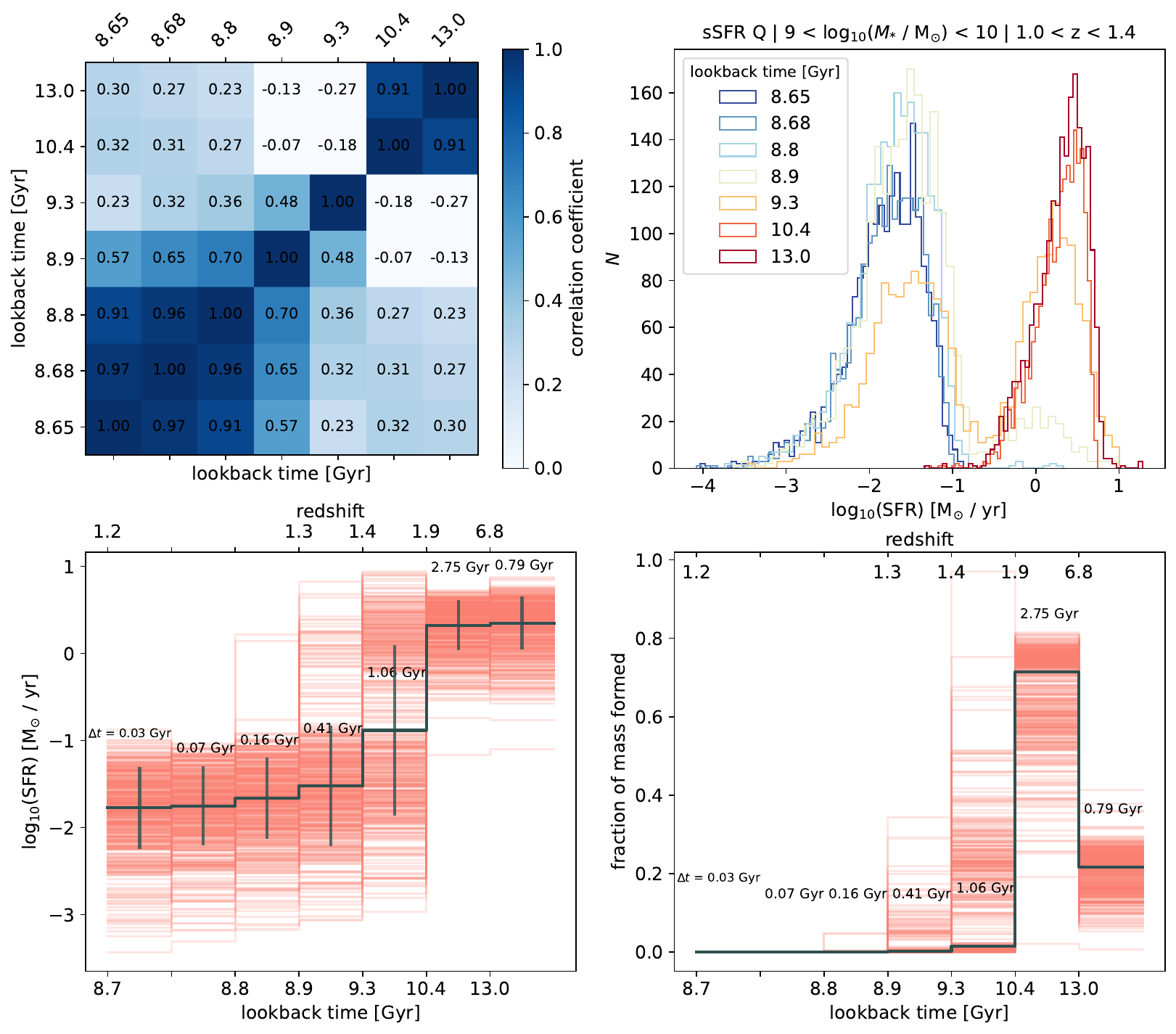}
    \caption{Same as Figure~\ref{fig:q_massive}, but for a lower stellar mass range of $9<\log_{10}(M_*/\mathrm{M}_\odot)<10$.}
    \label{fig:q_lowmass}
\end{figure*}

In this Appendix, we show the SFHs of \pc\ galaxies with lower stellar masses than those in Section \ref{sec:sfh}. Figures \ref{fig:sf_lowmass} and \ref{fig:q_lowmass} show, respectively, the SFHs of SF and Q galaxies at $z\simeq1.2$ with stellar masses $10^{9}$--$10^{10}\,\mathrm{M}_\odot$. For the same time intervals (400~Myr vs.\ 1~Gyr ago) discussed for the high-mass galaxies, the lower mass population shows slightly higher correlation for the subpopulations: $r = 0.81$ (SF) and $r = 0.48$ (Q). The SF subpopulation shows qualitatively the same trends as reported for the higher mass scale population in Section \ref{sec:sfh}. However, more quantitively, we see that the lower-mass galaxies reach a point where the correlation of their past SFR with its most recent value has dropped to $r<0.5$ at a shorter lookback time ($\sim 300$ Myr) compared to their higher mass counterparts (see top left panel of Figure \ref{fig:sf_massive}, $\sim 700$ Myr). Further, while the earliest and latest epochs in the more massive galaxies' SFH become entirely decorrelated over their history, for the lower mass subpopulation their SFH remains somewhat correlated over their entire lifespan. 

Some minor differences can be seen regarding the quenching transition in the Q subpopulation with respect to their more massive counterparts. Comparing the upper left panels of Figures \ref{fig:q_massive} and \ref{fig:q_lowmass}, the correlation structure in the lower-mass population is indicative of a more extended and complex quenching phase, extending over $\sim1.5$~Gyr in the fourth and fifth bins of the SFH. This is also clearly visible in the other panels of Figure \ref{fig:q_lowmass}, with two SFH bins showing bimodal SFR distributions in the upper right and lower left panels. We moreover see in the lower right panel of Figure \ref{fig:q_lowmass} that there are a non-negligible fraction of Q galaxies that formed as much as $\sim15$ percent of their total mass as recently as $\sim200$--700~Myr into their histories.

% Don't change these lines
\bsp	% typesetting comment
\label{lastpage}
\end{document}